\documentclass[aps,prl,reprint,superscriptaddress,amsmath,amssymb]{revtex4-2}

\usepackage{graphicx}
\usepackage{color}
\usepackage[percent]{overpic} 

\usepackage{mathtools}  
\usepackage{amssymb}
\usepackage{bm}
\usepackage{mathrsfs}
\usepackage{braket}
\usepackage{dsfont}
\usepackage{extarrows}
\usepackage{cancel}
\usepackage{xcolor}
\usepackage{xeCJK}

\usepackage[normalem]{ulem} 
\usepackage{enumitem}

\definecolor{xmorange}{RGB}{230,120,20} 

\usepackage[
  colorlinks,
  linkcolor=blue,
  citecolor=blue,
  urlcolor=blue,
  breaklinks=true
]{hyperref}

\begin{document}

\title{Heralded Free-Electron Writing of the Most Subradiant State in an Atomic Array}
\author{Tong Shen (沈彤)}
\affiliation{School of Physics and Technology, Wuhan University, Wuhan 430072, China}

\author{Zhexin Zhao}
\email{zhexin.zhao@fau.de}
\affiliation{Department of Physics, Friedrich-Alexander University (FAU) Erlangen-Nürnberg, Staudtstra{\ss}e 1, 91058 Erlangen, Germany}

\author{Meng Xiao}
\email{phmxiao@whu.edu.cn}
\affiliation{School of Physics and Technology, Wuhan University, Wuhan 430072, China}
\affiliation{Wuhan Institute of Quantum Technology, Wuhan 430206, China}
\begin{abstract}
The most subradiant eigenstate of a finite subwavelength atomic chain {in free space}, protected by strongly suppressed radiative decay, offers a powerful resource for {photon} storage, quantum sensing, and many-body quantum optics. Yet its optical preparation is hindered by the simultaneous need to match a wave vector outside the light cone and a nonuniform envelope. Here, we show that a free electron can overcome these constraints: its velocity sets the imprinted wave vector, while the trajectory {of the diffracting wave packet} shapes the excitation envelope. 
This simultaneous momentum and envelope matching enables heralded preparation with near-unity conditional fidelity (\(F>99.5\%\)) {even in a deeply subwavelength regime that is difficult to access with propagating free-space photons}. We further show that a path-superposed free electron can excite an antisymmetric state in two closely spaced parallel chains, {whose interchain destructive interference yields stronger subradiance than a single chain with the same total number of atoms. These results establish free electrons as quantum writers for collective excitations that are difficult to access with propagating optical fields.}

\end{abstract}

\maketitle

\textit{Introduction.}—Recent advances in ordered atomic arrays~\cite{endresAtombyatomAssemblyDefectfree2016,kaufmanQuantumScienceOptical2021,manetschTweezerArray61002025} have established quantum emitters with subwavelength spacing as a powerful platform for engineering cooperative optical responses~\cite{ruiSubradiantOpticalMirror2020,srakaewSubwavelengthAtomicArray2023,seubertTweezerAssistedSubwavelengthPositioning2025}. Subradiant collective states in such systems, arising from destructive interference in radiative emission, exhibit strongly suppressed decay and narrow linewidths, making them promising for {photon} storage~\cite{asenjo-garciaExponentialImprovementPhoton2017,manzoniOptimizationPhotonStorage2018,ferioliStorageReleaseSubradiant2021,ballantineQuantumSinglePhotonControl2021}, quantum sensing~\cite{ostermannProtectedStateEnhanced2013,facchinettiInteractionLightPlanar2018,zafra-bonoSubradiantCollectiveStates2025b}, and many-body quantum optics~\cite{perczelTopologicalQuantumOptics2017, douglasManyBodySuperSubradiance2026}. Yet the same radiative darkness that protects these states also makes them intrinsically difficult to address direct{ly} and write selectively. 
{In free space, existing optical strategies to access subradiant states include Stark- or Zeeman-assisted control}~\cite{plankensteinerSelectiveProtectedState2015,rubies-bigordaPhotonControlCoherent2022,fayardOpticalControlCollective2023} {and two-photon processes}~\cite{heAtomicSpinwaveControl2020a,rusconiExploitingPhotonicNonlinearity2021,cechDispersionlessSubradiantPhoton2023}. {These protocols typically rely on controlled Stark or Zeeman shifts, auxiliary atomic levels, or additional optical couplings.}
In the minimal setting of a finite one-dimensional chain of two-level atoms {with spacing $a<\lambda_0/2$, where $\lambda_0$ is the resonant transition wavelength}, the encoding challenge becomes particularly transparent: the most subradiant {single-excitation} state combines a Brillouin-zone-edge wave vector outside the light cone with a nonuniform spatial envelope~\cite{asenjo-garciaExponentialImprovementPhoton2017,zhangTheorySubradiantStates2019,zhangSubradiantEmissionRegular2020a}. Direct writing of such a state in a two-level atomic chain therefore requires simultaneous matching of the large wave vector and the spatial envelope, a task not naturally achievable with optical fields.

Free electron wave packets offer a complementary route to both controls and further abilities. Recent developments in free-electron quantum optics~\cite{garciadeabajoRoadmapQuantumNanophotonics2025,ruimyFreeelectronQuantumOptics2025} have shown that electron wave packets can excite discrete quantum systems~\cite{goverFreeElectronBoundElectronResonant2020,garciadeabajoCompleteExcitationDiscrete2022a, konecnaEntanglingFreeElectrons2022, gorlachDoublesuperradiantCathodoluminescence2024} and mediate quantum correlations~\cite{zhaoQuantumEntanglementModulation2021,kfirEntanglementsElectronsCavity2019b,rotunnoOneDimensionalGhostImaging2023,henkeProbingElectronphotonEntanglement2025,arendElectronsHeraldNonclassical2025}. {Unlike free-space photons, a free electron can transfer a longitudinal momentum tunable by its velocity, allowing access to the Brillouin-zone-edge wave vector even for deeply subwavelength atomic chains.}
At the same time, {the distance-dependent near-field electron--atom interaction~\cite{garciadeabajoOpticalExcitationsElectron2010} converts the electron trajectory and transverse profile into a controllable excitation envelope.} In addition, modern electron optics allows electron beams to be positioned and steered with nanometer-scale precision, providing the spatial resolution needed for selective coupling to atomic structures well below the optical diffraction limit~\cite{batsonSubangstromResolutionUsing2002,garciadeabajoProbingPhotonicLocal2008,polmanElectronbeamSpectroscopyNanophotonics2019a,auadMeVElectronSpectromicroscopy2023}. However, these capabilities have not yet been exploited for selective excitation of ordered atomic arrays.

In this Letter, we establish a free-electron--atomic-array interaction scheme for targeted writing of {the most} subradiant collective state, in which the electron wave packet is engineered in both momentum and trajectory to encode the desired eigenmode. The electron velocity provides momentum selectivity, while a uniform DC electric field steers the phase-matched electron wave packet along an optimized parabolic trajectory, thereby shaping the excitation profile to match the target subradiant state. This protocol allows the preparation fidelity to {surpass} the long-chain uniform-coupling bound of $8/\pi^2$. As a result, the most subradiant state can be prepared with near-unity conditional fidelity. This remains true {even at a deeply subwavelength spacing \(a=\lambda_0/20\)}. Successful preparation is heralded~\cite{feistCavitymediatedElectronphotonPairs2022, huangElectronPhotonQuantumState2023, arendElectronsHeraldNonclassical2025} by detecting an electron energy-loss event using electron energy-loss spectroscopy (EELS)~\cite{nazarovMultipoleSurfaceplasmonexcitationEnhancement1999,zewailFourDimensionalElectronMicroscopy2010,nazarovProbingMesoscopicCrystals2017,nazarovRoleKinematicsProbing2016}. We further extend the protocol to two parallel chains, where a path-superposed electron excites an antisymmetric collective state whose interchain destructive interference enables stronger subradiance {than a single chain with the same total number of atoms}.

\begin{figure}[t]
  \centering
  \includegraphics[width=\linewidth]{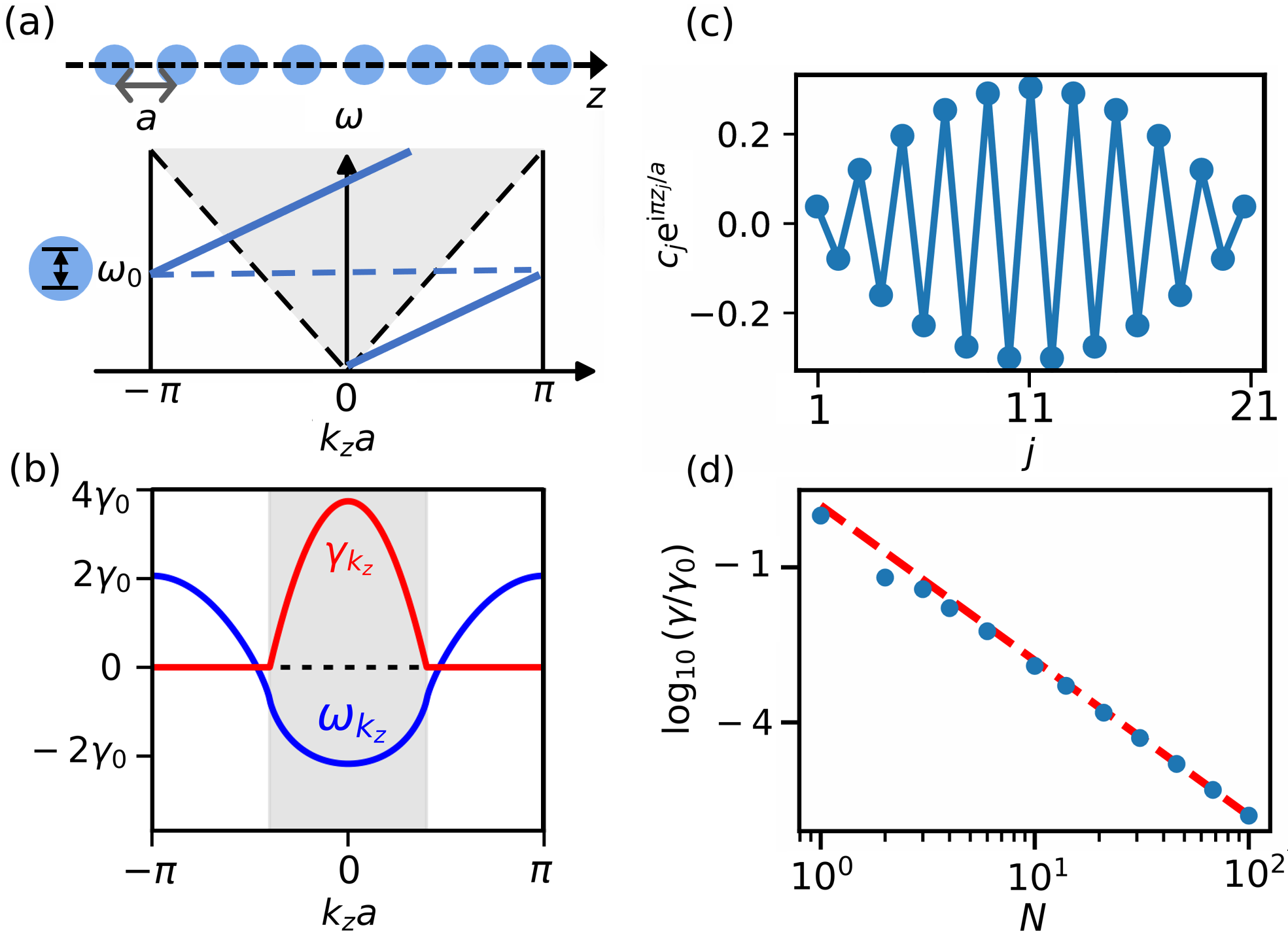}
  \caption{(a) Upper panel: Ordered atomic chain with spacing $a$ along the $z$ direction. Lower panel: Comparison between the free-space photon dispersion (black dashed lines) and the free-electron dispersion in the non-recoil approximation (blue solid line). The gray shaded area indicates the light cone, and $\omega_0$ is the transition frequency of each atom (modeled as a two-level system). (b) Collective frequency shift $\omega_{k_z}$ and decay rate $\gamma_{k_z}$ of the Bloch modes for an infinite atomic chain, where $\gamma_0$ is the single-atom decay rate and $a=\lambda_0/5$. The decay rate vanishes outside the light cone (gray region). (c) Coefficients $c_j\exp\left({{\text{i}\pi z_j}/{a}}\right)$ of the most subradiant state in a finite chain with 21 atoms. (d) Normalized minimum decay rate $\gamma_{\min}/\gamma_0$ versus the atom number $N$, where the red dashed line shows the power-law $\sim N^{-3}$.}
  \label{fig:1}
\end{figure}

{
\textit{One-dimensional chain and phase-matching.}—We first characterize the collective modes of a one-dimensional atomic chain and identify the target most subradiant state. We consider \(N\) identical two-level atoms in free space, arranged in a chain with spacing \(a\) [Fig.~\hyperref[fig:1]{1(a)}]. Within the Born--Markov approximation, their collective eigenmodes are described by the effective non-Hermitian Hamiltonian~\cite{asenjo-garciaExponentialImprovementPhoton2017} (see Sec.~I of the Supplemental Material (SM)~\cite{supplemental})
\begin{equation}
\label{eq:Heff}
    \mathcal H_{\text{eff}}=-\mu_0\omega_0^2\sum_{i,j=1}^N\mathbf{d}_i^{*}\cdot\mathbf G(\mathbf r_i,\mathbf r_j,\omega_0)\cdot\mathbf{d}_j\sigma^+_i\sigma_j^-,
\end{equation}
where \(\omega_0\) is the transition frequency between the excited state \(\ket{e}\) and the ground state \(\ket{g}\), \(\sigma_i^+=\ket{e_i}\bra{g_i}\) is the raising operator of the $i$th atom, \(\mathbf d_i=-\mathrm{e}\,\boldsymbol{l}_i=-\mathrm{e}\braket{e_i|\mathbf r|g_i}\) is the transition dipole moment, and \(\mathbf G\) is the free-space dyadic Green's tensor. Throughout this Letter, we take the chain axis to be along \(z\) and use \(z\)-polarized transition dipoles as a representative case, while restricting the dynamics to the single-excitation subspace. Other dipole orientations can be treated within the same formalism by replacing the corresponding Green-tensor and electron--atom coupling components.

For an infinite chain, the non-Hermitian Hamiltonian is diagonalized by Bloch-like eigenstates \(\sum_j\mathrm{e}^{\mathrm{i}k_zz_j}\sigma^+_j\ket{G}\) with complex eigenvalues \(\hbar(\omega_{k_z}-\mathrm{i}{\gamma_{k_z}}/{2})\). Here \(k_z\) is the Bloch wave vector, \(z_j\) is the position of the \(j\)th atom, and \(\ket{G}\) is the collective ground state. \(\omega_{k_z}\) and \(\gamma_{k_z}\) are the collective frequency shift and collective decay rate, respectively, as illustrated for $a/\lambda_0=1/5$ in Fig.~\hyperref[fig:1]{1(b)}. For \(a<\lambda_0/2\), the decay rate vanishes when \(|k_z|>k_0\), with $k_0=\omega_0/c$. A free-space photon incident on the atomic array can provide only a longitudinal wave-vector component satisfying \(|k_z|\leq k_0\), as indicated by the gray shaded region in Fig.~\hyperref[fig:1]{1(a)}. By contrast, for an electron propagating along the \(z\) direction with central momentum \(p_0\), the free-electron dispersion can be linearized within the nonrecoil approximation as \(E(p_z)\simeq E_0+v_0(p_z-p_0)\), where \(E_0\) and \(v_0\) are the corresponding central energy and velocity. When the atomic chain absorbs an energy quantum \(\hbar\omega_0\), energy conservation together with this linearized dispersion gives a longitudinal momentum transfer \(\hbar q\), with \(q=\omega_0/v_0\). This relation is illustrated by the solid blue line in Fig.~\hyperref[fig:1]{1(a)}, whose slope is set by the electron velocity \(v_0\). 

For a finite chain, direct diagonalization of Eq.~\eqref{eq:Heff} identifies the most subradiant state as the eigenstate with the smallest decay rate. This state can be written as
\begin{equation}
    \ket{\psi_{\text{sub}}}=\sum_j c_j \mathrm{e}^{\mathrm{i}\pi z_j/a}\mathrm{e}^{\mathrm{i}\delta \varphi_j}\sigma_j^+\ket{G},
\end{equation}
where \(c_j\) describes the nonuniform amplitude envelope and \(\delta\varphi_j\) denotes a small phase correction to the dominant staggered phase (see the SM Sec. I~\cite{supplemental}). The distribution of \(c_j\mathrm{e}^{\mathrm{i}\pi z_j/a}\) for \(N=21\) is shown in Fig.~\hyperref[fig:1]{1(c)}. This nonuniform envelope is the key feature that distinguishes the most subradiant state from a uniformly weighted Bloch wave in the infinite chain case. Meanwhile, the minimum decay rate decreases approximately as \(\gamma_{\min}\sim N^{-3}\)~\cite{asenjo-garciaExponentialImprovementPhoton2017}, as shown in Fig.~\hyperref[fig:1]{1(d)}. To imprint the Brillouin-zone-edge wave vector $k_z=\pi/a$ of the target state with a free electron, we choose \(v_0=v_n\) such that \(\omega_0/v_n=(2n+1)\pi/a\), where \(n\) denotes the phase-matching order. As a deeply subwavelength example, when \(a=\lambda_0/20\), the Brillouin-zone edge lies at \(\pi/a=10k_0\), well outside the momentum range of propagating optical fields. A free electron can nevertheless reach it by choosing \(v_n=c/[10(2n+1)]\). However, this phase-matching picture alone provides only a kinematic guide; we next formulate the full electron--chain scattering problem.
}

\begin{figure}[h]
  \centering
  \includegraphics[width=\linewidth]{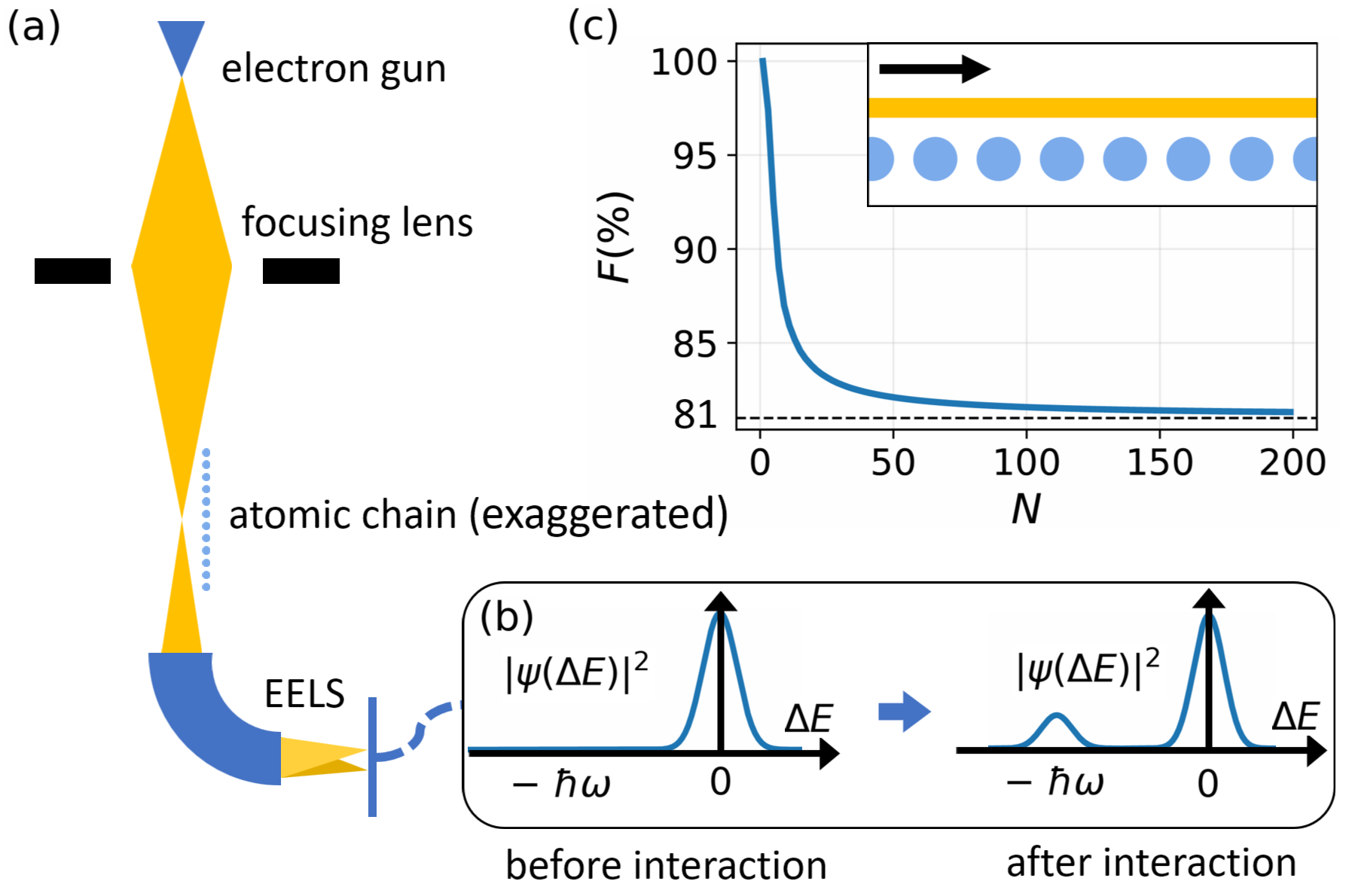}
  \caption{Free-electron heralding protocol and straight-trajectory benchmark.
    (a) Schematic of the EELS-based setup. A quasi-monochromatic electron is emitted from an electron gun, focused onto the atomic chain by an electromagnetic lens, and analyzed after the interaction by EELS.
    (b) Electron energy spectrum before and after the interaction. 
    (c) Conditional fidelity $F$ for preparing the most subradiant state using a diffraction-free, well-focused straight electron trajectory, illustrated in the inset. The fidelity approaches the $8/\pi^2\approx 81\%$ limit for large $N$.
    }
  \label{fig:2}
\end{figure}

{\textit{Electron--chain scattering model.}}---Our model is sketched in Fig.~\hyperref[fig:2]{2(a)}. A quasi-monochromatic electron with spectral width \(\Delta E_{\text{FWHM}} \ll \hbar\omega_0\), whose initial energy spectrum is shown in the left panel of Fig.~\hyperref[fig:2]{2(b)}, is emitted from an electron gun and focused by an electromagnetic lens onto the atomic chain. After the interaction, the electron energy spectrum is analyzed by EELS. The detection of an energy-loss event \(\hbar\omega_0\), appearing as the loss peak in the right panel of Fig.~\hyperref[fig:2]{2(b)}, heralds that a single collective excitation has been written into the atomic chain.

The total Hamiltonian reads
\begin{equation}
    H=H_{\text{free}}+H_{\text{atom}}+H_{\text{int}},
\end{equation}
where \(H_{\text{free}}=E_0+v_0(p_z-p_0)\) is the free-electron Hamiltonian and the interaction term \(H_{\text{int}}\) is the instantaneous Coulomb interaction. Owing to the weak electron–atom interaction, we treat it perturbatively. To the first order in the Magnus expansion, the scattering operator is (Sec.~II of the SM~\cite{supplemental})
\begin{equation}
    S=\exp\left[-\mathrm{i}\left(gb S_++\text{h.c.}\right)\right],
\end{equation}
where \(b=\exp\left(-\mathrm{i}\omega_0z/v_0\right)\) is the electron ladder operator~\cite{feistQuantumCoherentOptical2015,kfirEntanglementsElectronsCavity2019b,ruimyAtomicResolutionQuantumMeasurements2021}, corresponding to an energy loss \(\hbar\omega_0\) and a longitudinal momentum change \(-\hbar\omega_0/v_0\), with \(z\) the electron position operator. The collective raising operator is
\begin{equation}
    S_+=\frac{1 }{g}\sum_{j}g_j\mathrm{e}^{\mathrm{i}\omega_0z_j/v_0}\sigma^+_j,
\end{equation}
where \(g=\sqrt{\sum_j|g_j|^2}\), \(g_j\) is the coupling amplitude between the electron and the \(j\)th atom, and \(|g|^2\) is the heralding probability in the weak-coupling regime.

Let the initial state be \(\ket{\Psi_i}=\ket{G}\otimes\ket{\psi_e}\), where $\ket{\psi_e}$ is the initial free-electron state. In the energy representation, \(|\braket{E|\psi_e}|^2\) is taken to satisfy a narrow Gaussian distribution with a width much smaller than \(\hbar\omega_0\). After the interaction, the final state becomes
\begin{equation}
    \ket{\Psi_f}=\left(1-\frac{|g|^2 }{2}\right)\ket{G}\otimes\ket{\psi_e}-\mathrm{i}gS_+\ket{G}\otimes b\ket{\psi_e}+O(|g|^3).
\end{equation}
Resolving the shifted electron state \(b\ket{\psi_e}\) in EELS is therefore sufficient to herald the atomic excitation \(S_+\ket{G}\). In the electron spectrum, this process appears as a loss peak at $\Delta E=E-E_0=-\hbar\omega_0$, as shown in the right panel of Fig.~\hyperref[fig:2]{2(b)}, where $E$ is the outgoing electron energy and $E_0$ is the incident central energy.

To show the imprinted wave vector and the necessity of non-straight trajectory, we first consider a diffraction-free, well-focused electron on a straight trajectory (Sec.~III of the SM~\cite{supplemental}), as sketched in the inset of Fig.~\hyperref[fig:2]{2(c)}. In this limit, the transverse distribution of the electron is approximated by a Dirac \(\delta\) function, as commonly assumed in free-electron--bound-electron interactions~\cite{goverFreeElectronBoundElectronResonant2020,zhaoQuantumEntanglementModulation2021,ruimyAtomicResolutionQuantumMeasurements2021}, and the coupling amplitude is uniform along the chain, {leading to $S_+=\frac{1}{\sqrt{N}}\sum_j\text{e}^{\text i \omega_0z_j/v_0}\sigma^+_j$,} so the prepared state is close to a uniformly weighted staggered state rather than the eigenstate in Fig.~\hyperref[fig:1]{1(c)}. The resulting conditional fidelity \(F=|\braket{\psi_{\text{sub}}|S_+|G}|^2\) is shown in Fig.~\hyperref[fig:2]{2(c)}. Even {though} the phase-matching condition is satisfied, the fidelity approaches the uniform-coupling bound \(8/\pi^2\) in the large-\(N\) limit (Sec.~III of the SM~\cite{supplemental}). 
Moreover, this result assumes an ideal diffraction-free electron beam approximation, {which is violated when the interaction length is long}. 

\begin{figure}[t]
  \centering
  \includegraphics[width=1\linewidth]{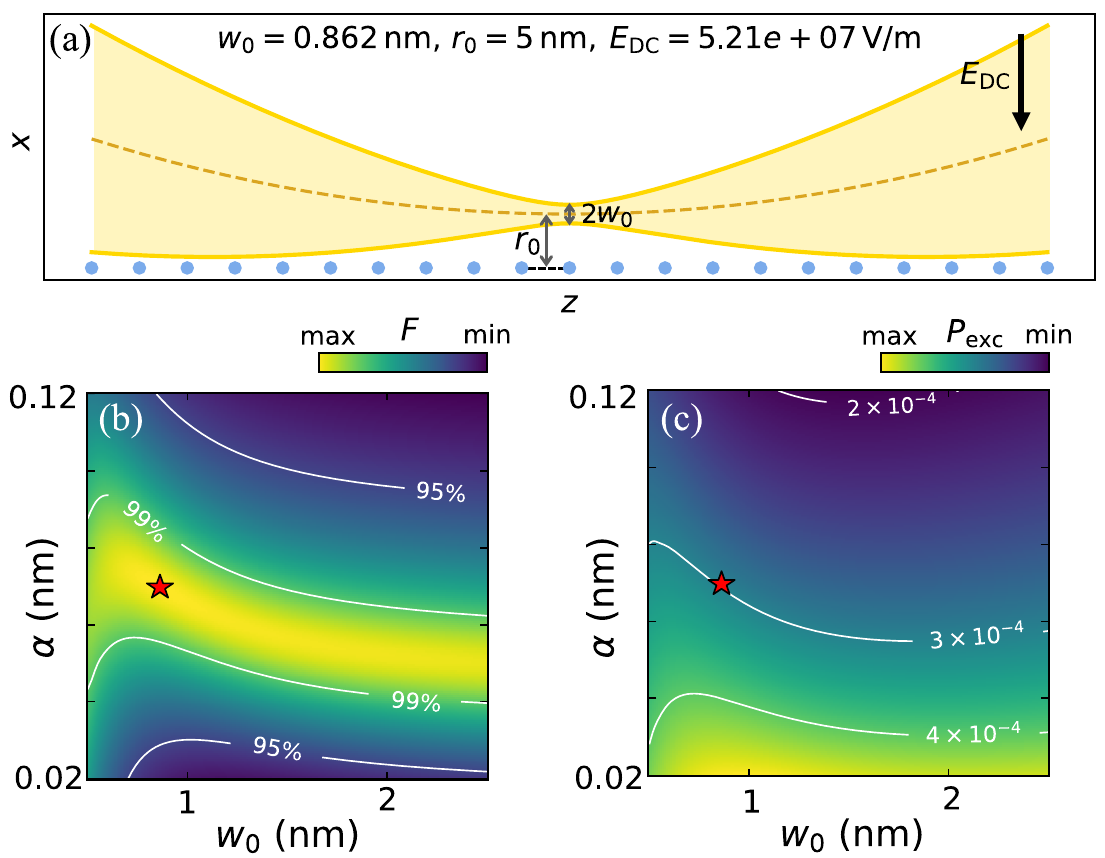}
  \caption{
        (a) Schematic of a diffracting Gaussian electron wave packet steered by a vertical DC electric field \(E_{\rm DC}\). {The dashed line marks the center of the electron wave packet, while the} yellow shaded region denotes \(r_{\rm mid}(z)\pm w(z)\). {The parameters correspond to the optimal point marked by the red stars in (b) and (c)}. {Other parameters are $N=21$, $\lambda_0=780$ nm, $a/\lambda_0=1/20$, $n=1$ (with electron velocity $v_0=c/30$), and $r_0=5$ nm.}
        (b, c) Conditional fidelity (b) and excitation probability (c) as functions of the beam waist \(w_0\) and trajectory curvature \(\alpha\). The red star indicates the maximum-fidelity point, with \(F_{\text{max}}=99.55\%\) and \(P_{\text{exc}}=2.97\times10^{-4}\).
}
  \label{fig:3}
\end{figure}

\textit{Diffracting parabolic trajectory.}—We choose a parabolic trajectory so that the electron passes closest to the chain center, where the coupling is strongest, and gradually moves away toward the edges. Such a trajectory is naturally generated by a uniform transverse DC electric field. For a homogeneous transverse DC field, the induced Stark shift is the same for all {two-level} atoms and is thus absorbed into \(\omega_0\). To describe this geometry, we generalize the electron--atom interaction framework to a wave packet with a varying transverse profile. We assume that the transverse profile is not modified by the weak interaction with the atomic chain. The coupling strength to an atom at longitudinal position \(z=z_j\) with a monochromatic electron is then (Sec.~II of the SM~\cite{supplemental})
\begin{equation}
    g_j=\frac{e^2}{4\pi\varepsilon_0\hbar v_0}\int\mathrm d^3\mathbf r\,|\phi_\perp(\mathbf r_\perp,z)|^2\mathrm e^{\text{i}\frac{\omega_0}{v_0}(z-z_j)}\frac{\boldsymbol l_j\cdot(\mathbf r-z_j\hat{z})}{|\mathbf r-z_j\hat{z}|^3},
\end{equation}
where \(\phi_{\perp}(\boldsymbol{r}_{\perp},z)\) is the transverse wave function {of the electron after considering diffraction and the external DC field}, and $\boldsymbol{l}_j$ is the transition length of the $j$th atom.
Then we model the electron as a Gaussian beam with waist \(w_0\) [see Fig.~\hyperref[fig:3]{3(a)}]. Under a vertical DC electric field \(E_{\rm DC}\), the transverse distribution remains Gaussian, while its center follows a parabolic trajectory~\cite{kudlisEvolutionTwistedElectron2026}. The beam center is taken as \(\mathbf r_{\mathrm{mid}}(z)=\left[r_0+\alpha\left(z/a\right)^2\right]\hat{x}\) [denoted by the dashed line in Fig.~\hyperref[fig:3]{3(a)}], where \(r_0\) is the minimum electron--chain distance and \(\alpha\) sets the curvature. The latter can be tuned by the external DC field as \(\alpha=eE_{\mathrm{DC}}a^2/(2m_ev_0^2)\). The chain is assumed to be symmetric about \(z=0\). The transverse probability distribution is
\begin{equation}
    |\phi_{\perp}(\boldsymbol{r}_{\perp},z)|^2=\frac{2 }{\pi}\frac{1}{w^2(z)}\exp\left[-\frac{2|\boldsymbol{r}_{\perp}-\mathbf r_{\mathrm{mid}}(z)|^2 }{w^2(z)}\right],
\end{equation}
where $w(z)=w_0\sqrt{1+(z/z_R)^2},z_R=\pi w_0^2/\lambda_e$, 
with \(z_R\) the Rayleigh length and \(\lambda_e\) the de Broglie wavelength of the electron. 

{

By tuning the beam waist \(w_0\), the trajectory curvature \(\alpha\), and selecting appropriate values of the phase-matching order \(n\) and of \(r_0\), our method can excite the most subradiant state with fidelity \(F>99.5\%\) over a wide range of \(N\) and \(a\) (\(N=20\text{--}100\) and \(a/\lambda_0=1/20\text{--}1/5\), see Sec.~V of the SM~\cite{supplemental}). The corresponding excitation probability ranges from \(10^{-6}\) to \(10^{-4}\). For a conventional EELS probe current of \(I\sim 10^2\,\mathrm{pA}\)~\cite{marinovaSTEMEELSInvestigationPlanar2021,roccaprioreDynamicSTEMEELSSingleatom}, this gives an appreciable heralding rate of \(R_{\mathrm{herald}}\sim10^{2}\text{--}10^{5}\,\mathrm{s}^{-1}\). The required external electric field is on the order of \(10^{6}\) to \(10^{7}\,\mathrm{V/m}\). As a deeply subwavelength example, in Fig.~\ref{fig:3} we show the case of \(N=21\) and \(a/\lambda_0=1/20\). Here, Fig.~\ref{fig:3}(a) shows the optimized scheme, while Figs.~\ref{fig:3}(b) and \ref{fig:3}(c) show the landscapes of \(F\) and \(P_{\mathrm{exc}}\) as functions of \(w_0\) and \(\alpha\), respectively. We use the \(D_2\) line of \({}^{87}\mathrm{Rb}\)~\cite{SteckRb87} as a representative numerical example: $\lambda_0=780$ nm and $l^{21}=0.158$ nm. Also, for an electron with characteristic spectral width $\Delta E_{\text{FWHM}}=0.1\text{ eV}$~\cite{mankos2012progress,lozano2018magnetically,lozano2025electron,tromp2023gun}, the fidelity change remains on the order of \(10^{-4}\), as discussed in Sec.~VI of the SM~\cite{supplemental}.
}



\begin{figure}[b]
  \centering
  \includegraphics[width=\linewidth]{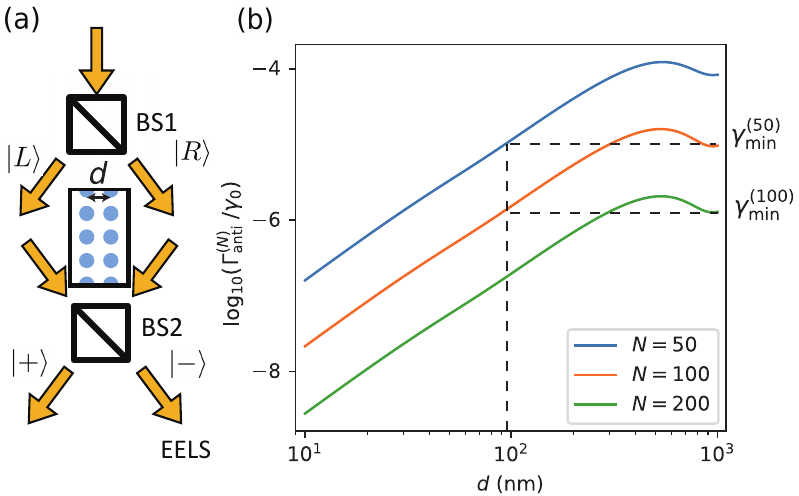}
  \caption{
  (a) Scheme for heralding an antisymmetric excitation in two parallel atomic chains. A path-superposed electron interacts with the two chains and is recombined before EELS detection. The inset defines the interchain spacing \(d\). 
  (b) Decay rate \(\Gamma^{(N)}_{\text{anti}}/\gamma_0\) of the antisymmetric state as a function of the interchain spacing \(d\) for several total atom numbers \(N\). All other unspecified parameters are the same as those in Fig.~\hyperref[fig:3]{3}. {The horizontal dashed lines indicate \(\gamma_{\min}^{(50)}\) and \(\gamma_{\min}^{(100)}\) for single chains, serving as references for the two-chain decay suppression. The vertical dashed line marks \(d=98~\mathrm{nm}\), used as a representative interchain spacing for this comparison.}
  }
  \label{fig:4}
\end{figure}

\textit{{Antisymmetric} subradiant states.}---The protocol above provides a building block for more complex collective states.  Free electrons can have de Broglie wavelengths \(h/p\) on the \AA{}ngstr\"om scale, enabling spatial resolution far beyond that of photons. 
Combined with their near-field dipole interaction with atoms, this allows nanometer-scale spatial selectivity in addressing atomic arrays~\cite{batsonSubangstromResolutionUsing2002,garciadeabajoProbingPhotonicLocal2008,polmanElectronbeamSpectroscopyNanophotonics2019a,auadMeVElectronSpectromicroscopy2023}.
As shown in Fig.~\hyperref[fig:4]{4(a)}, we consider two identical chains, denoted by \(L\) and \(R\), each containing \(N/2\) atoms and separated by a distance \(d\). 
The first beam splitter~\cite{shindoAdvancedElectronHolography2017,johnsonScanningTwogratingFree2021,johnsonInelasticMachZehnderInterferometry2022} prepares the electron in the path superposition \((\ket{L}+\text{i}\ket{R})/\sqrt{2}\), with the two paths aligned with the two chains. 
After the interaction, the one-energy-quantum-loss component is proportional to \(g_L b\ket{L}\ket{1_L}+\text{i}g_R b\ket{R}\ket{1_R}\), where \(b\ket{L}\) and $b\ket{R}$ denote the electron path states after losing one quantum \(\hbar\omega_0\), and \(\ket{1_{L,R}}=S_{L,R,+}\ket{G_LG_R}\) are the collective excitations written into the two chains. 
For balanced coupling strengths \(g_L=g_R\), the second beam splitter erases the path information, so that EELS detection in the two output ports projects the atomic chains onto \(\ket{\psi_{\pm}}=(\ket{1_L}\pm\ket{1_R})/\sqrt{2}\), up to an overall phase. 
The \(\ket{-}\) port therefore prepares the antisymmetric excitation
\begin{equation}
    \ket{\psi_A}
    =\frac{1}{\sqrt{2}}\sum_{j=1}^{N/2}c_j
    \mathrm{e}^{\text{i}\pi z_j/a}\text{e}^{\text i\delta\varphi_j}
    \left(\sigma_{Lj}^{+}-\sigma_{Rj}^{+}\right)\ket{G_LG_R}.
\end{equation} 
This antisymmetric superposition can further suppress collective radiation at small \(d\). We note, however, that reducing the interchain separation can also introduce electron-path crosstalk, which may lower the preparation fidelity. Nevertheless, with suitable parameters, the conditional fidelity can still exceed \(99\%\); a representative optimization including this crosstalk is provided in Sec.~VIII of the SM~\cite{supplemental}.

We now proceed to evaluate the radiative decay of the antisymmetric state. Let \(\gamma^{(N)}_{\min}\) and \(\Gamma^{(N)}_{\text{anti}}\) denote the decay rates of the most subradiant state of an \(N\)-atom single chain and of the antisymmetric state of the two parallel chains with the same total number \(N\) of atoms, respectively. Figure~\hyperref[fig:4]{4(b)} shows \(\Gamma^{(N)}_{\text{anti}}\), obtained by diagonalizing the effective Hamiltonian in Eq.~\eqref{eq:Heff} within the single-excitation subspace. 
The relation between \(\gamma^{(N/2)}_{\text{min}}\) and \(\Gamma^{(N)}_{\text{anti}}\) is well captured by the two-dipole interference factor (Sec.~VII of the SM~\cite{supplemental}),
\begin{equation}
    \frac{\Gamma^{(N)}_{\text{anti}}}{\gamma_{\min}^{(N/2)}}\approx
    1-\frac{3}{2}
    \left[
    \frac{\sin(k_0d)}{k_0d}
    +\frac{\cos(k_0d)}{(k_0d)^2}
    -\frac{\sin(k_0d)}{(k_0d)^3}
    \right].
\end{equation}
In the limit \(d\rightarrow+\infty\), the two chains become independent and \(\Gamma^{(N)}_{\text{anti}}\rightarrow\gamma_{\min}^{(N/2)}\). 
At finite \(d\), however, destructive interference between the two chains further suppresses radiative decay, allowing the collective decay rate to fall below the single-chain value. 
For \(k_0d\ll1\), the decay-rate ratio scales as \(\Gamma^{(N)}_{\text{anti}}/\gamma_{\text{min}}^{(N/2)}\approx(k_0d)^2/5\). {
The \(N^{-3}\) scaling in Fig.~\hyperref[fig:1]{1(d)} implies \(\gamma^{(N)}_{\text{min}}/\gamma^{(N/2)}_{\text{min}}\approx 1/8\) for a single chain. 
Combining this scaling with the interchain interference factor shows that an interchain spacing satisfying \(k_0d<\sqrt{5/8}\), or \(d<98~\mathrm{nm}\) for the \(D_2\) line of \({}^{87}\mathrm{Rb}\), enables the two-chain configuration to suppress radiative decay more strongly than a single chain with the same total number of atoms (as indicated by the dashed line in Fig.~\hyperref[fig:4]{4(b)}).}

\textit{Conclusions.}—We have proposed a free-electron heralding protocol for preparing the most subradiant collective states of ordered atomic chains. The key ingredients are the tunable longitudinal momentum of the free electron, which matches the staggered phase of the target state, and an optimized parabolic trajectory, which shapes the coupling profile to match the nonuniform finite-chain envelope. Together, these controls enable near-unity conditional fidelity even in the deeply subwavelength regime. We further showed that path-superposed free electrons can herald chain-entangled states in two parallel atomic chains, where interchain destructive interference provides an additional suppression of the radiative decay rate. {Although we have focused on atomic arrays, the protocol can in principle be extended to other ordered quantum emitters, including quantum dots~\cite{tiranovCollectiveSuperSubradiant2023,kimCavitymediatedCollectiveEmission2025} and molecules~\cite{trebbiaTailoringSuperradiantSubradiant2022,langeSuperradiantSubradiantStates2024}}. {Our results identify free electrons as new tools for preparing long-lived collective matter states that are difficult to access with propagating optical fields. Looking ahead, further control of free-electron wave packets may enable the selective writing of other collective eigenmodes and even arbitrary collective excitations in atomic arrays.
}

\textit{Acknowledgments.}—We thank Prof. Peter Baum, Dr. Yiqi Fang, Prof. Yiming Pan, Prof. Ana Asenjo-Garcia, and Prof. Yu-Xiang Zhang for helpful discussions and suggestions. M.X. acknowledges support from the Key Research and Development Program of the Ministry of Science and Technology (Grant No. 2024YFB2808200), the National Natural Science Foundation of China (Grant Nos. 12321161645, 12334015, and 12274332), and the National Key Research and Development Program of China (Grant No. 2022YFA1404900). Z.Z. acknowledges support from the Alexander von Humboldt Foundation through a Postdoctoral Fellowship.

\textit{Data Availability.}—The numerical data and code that support the findings of this study are available from the corresponding authors upon reasonable request.

\bibliographystyle{apsrev4-2}
\bibliography{main}
\end{document}


\begin{center}
{\large \textbf{Supplemental Material for\\
Heralded Free-Electron Writing of the Most Subradiant State in an Atomic Array}}

\vspace{1em}

Tong Shen$^{1}$ (沈彤), Zhexin Zhao$^{2,*}$, and Meng Xiao$^{1,3,\dagger}$

\vspace{0.5em}

{\small
$^{1}$\textit{School of Physics and Technology, Wuhan University, Wuhan 430072, China}\\
$^{2}$\textit{Department of Physics, Friedrich-Alexander University (FAU) Erlangen-Nürnberg,
Staudtstraße 1, 91058 Erlangen, Germany}\\
$^{3}$\textit{Wuhan Institute of Quantum Technology, Wuhan 430206, China}\\
$^{*}$zhexin.zhao@fau.de\\
$^{\dagger}$phmxiao@whu.edu.cn
}
\end{center}
\vspace{1em}

\tableofcontents
\newpage
\section{I. The dyadic Green's tensor and collective eigenmodes}

{This section follows the standard Green-tensor formulation of collective radiative coupling in ordered atomic arrays~\cite{asenjo-garciaExponentialImprovementPhoton2017}.}
We first recall how the dyadic Green's tensor enters the interaction between a set of dipoles in free space. Consider \(N\) classical oscillating dipoles located at positions \(\mathbf r_j\), with dipole moments \(\mathbf p_j(\omega)\) at frequency \(\omega\). The electric field generated by these dipoles satisfies the frequency-domain wave equation
\begin{equation}
    \nabla\times\nabla\times\mathbf E(\mathbf r,\omega)-\frac{\omega^2}{c^2}\mathbf E(\mathbf r,\omega)=\mu_0\omega^2\sum_{j=1}^{N}\mathbf p_j(\omega)\delta(\mathbf r-\mathbf r_j).
\end{equation}
To solve this equation, we introduce the free-space dyadic Green's tensor \(\mathbf G(\mathbf r,\mathbf r',\omega)\), defined by
\begin{equation}
    \nabla\times\nabla\times\mathbf G(\mathbf r,\mathbf r',\omega)-\frac{\omega^2}{c^2}\mathbf G(\mathbf r,\mathbf r',\omega)=\delta(\mathbf r-\mathbf r')\mathds 1.
\end{equation}
The electric field can then be written as
\begin{equation}
    \mathbf E(\mathbf r,\omega)=\mathbf E_{\mathrm{in}}(\mathbf r,\omega)+\mu_0\omega^2\sum_{j=1}^{N}\mathbf G(\mathbf r,\mathbf r_j,\omega)\cdot\mathbf p_j(\omega),
\end{equation}
where \(\mathbf E_{\mathrm{in}}\) is the incident field. Thus, the Green's tensor $\mathbf G(\mathbf r,\mathbf r_j,\omega)$ gives the field at \(\mathbf r\) generated by a point dipole at \(\mathbf r_j\).

In free space, translational invariance gives \(\mathbf G(\mathbf r,\mathbf r',\omega)=\mathbf G(\mathbf r-\mathbf r',\omega)\). The explicit solution is
\begin{equation}
    \mathbf G(\mathbf r,\omega)=\frac{\mathrm e^{\mathrm i k r}}{4\pi k^2r^3}\left[\left(k^2r^2+\mathrm i kr-1\right)\mathds 1+\left(-k^2r^2-3\mathrm i kr+3\right)\frac{\mathbf r\otimes\mathbf r}{r^2}\right],
\end{equation}
where \(k=\omega/c\), and in this equation \(\mathbf r\) denotes the displacement \(\mathbf r-\mathbf r'\), with \(r=|\mathbf r|\). This expression contains both the coherent near-field dipole--dipole interaction and the radiative far-field contribution. Its real part gives the coherent exchange interaction between dipoles, while its imaginary part gives the radiative decay and collective emission.

The same Green-tensor structure can be used to describe quantum dipoles. In the Markov approximation, the external electromagnetic field can be treated as a reservoir, and the remaining dynamics of the atomic degrees of freedom is governed by a master equation involving only atomic operators. For an ensemble of two-level atoms with ground state \(\ket{g}\), excited state \(\ket{e}\), transition frequency \(\omega_0\), and transition dipole moment
\begin{equation}
    \mathbf d_i=-e\boldsymbol l_i=-e\braket{e_i|\mathbf r|g_i},
\end{equation}
the reduced density matrix of the atoms obeys
\begin{equation}
    \frac{\mathrm d\rho}{\mathrm dt}=-\frac{\mathrm i}{\hbar}[\mathcal H,\rho]+\mathcal L[\rho].
\end{equation}
Here
\begin{equation}
    \mathcal H=\hbar\omega_0\sum_{i=1}^{N}\sigma_i^+\sigma_i^-+\hbar\sum_{i,j=1}^{N}J^{ij}\sigma_i^+\sigma_j^-,
\end{equation}
and
\begin{equation}
    \mathcal L[\rho]=\sum_{i,j=1}^{N}\frac{\Gamma^{ij}}{2}\left(2\sigma_j^-\rho\sigma_i^+-\sigma_i^+\sigma_j^-\rho-\rho\sigma_i^+\sigma_j^-\right).
\end{equation}
The atomic raising and lowering operators are \(\sigma_i^+=\ket{e_i}\bra{g_i}\) and \(\sigma_i^-=(\sigma_i^+)^\dagger\). The coherent and dissipative coupling rates are determined by the real and imaginary parts of the dyadic Green's tensor,
\begin{equation}
    J^{ij}=-\frac{\mu_0\omega_0^2}{\hbar}\mathbf d_i^*\cdot\mathrm{Re}\,\mathbf G(\mathbf r_i,\mathbf r_j,\omega_0)\cdot\mathbf d_j,
\end{equation}
and
\begin{equation}
    \Gamma^{ij}=\frac{2\mu_0\omega_0^2}{\hbar}\mathbf d_i^*\cdot\mathrm{Im}\,\mathbf G(\mathbf r_i,\mathbf r_j,\omega_0)\cdot\mathbf d_j.
\end{equation}
For \(i=j\), \(\Gamma^{ii}=\gamma_0\) is the single-atom spontaneous-emission rate,
\begin{equation}
    \gamma_0=\frac{\omega_0^3|\mathbf d_i|^2}{3\pi\hbar\varepsilon_0c^3}.
\end{equation}
The divergent real part of the self-interaction is absorbed into the renormalized transition frequency \(\omega_0\).

The Lindblad master equation can also be written in the quantum-jump form~\cite{dalibardWaveFunctionApproach1992,dumMonteCarloSimulation1992}. To this end, we separate the no-jump evolution from the recycling term and rewrite the master equation as
\begin{equation}\label{eq:jump_master}
    \frac{\mathrm d\rho}{\mathrm dt}=-\frac{\mathrm i}{\hbar}\left(\mathcal H_{\mathrm{eff}}\rho-\rho\mathcal H_{\mathrm{eff}}^{\dagger}\right)+\sum_{i,j}\Gamma^{ij}\sigma_j^-\rho\sigma_i^+.
\end{equation}
Here the effective non-Hermitian Hamiltonian is
\begin{equation}
    \mathcal H_{\mathrm{eff}}=\mathcal H-\frac{\mathrm i\hbar}{2}\sum_{i,j=1}^{N}\Gamma^{ij}\sigma_i^+\sigma_j^-.
\end{equation}
Substituting the expressions for \(\mathcal H\), \(J^{ij}\), and \(\Gamma^{ij}\), and absorbing the single-atom Lamb shift into the renormalized transition frequency \(\omega_0\), one obtains (where the trivial \(\hbar\omega_0\) term has been omitted by working in the rotating frame)
\begin{equation}
    \mathcal H_{\mathrm{eff}}=-\mu_0\omega_0^2\sum_{i,j=1}^{N}\mathbf d_i^*\cdot\mathbf G(\mathbf r_i,\mathbf r_j,\omega_0)\cdot\mathbf d_j\,\sigma_i^+\sigma_j^-.
\end{equation}
The first term in Eq.~\eqref{eq:jump_master} describes the conditional evolution of the atomic state in the absence of a detected photon emission event. Because \(\mathcal H_{\mathrm{eff}}\) is non-Hermitian, this no-jump evolution is not norm preserving; the decrease of the norm gives the probability that a quantum jump has occurred. The second term, \(\sum_{i,j}\Gamma^{ij}\sigma_j^-\rho\sigma_i^+\), is the recycling term associated with stochastic photon emission. In the single-excitation subspace considered here, such a jump removes the atomic excitation and transfers the system to the collective ground state \(\ket{G}\).

To make this equivalence explicit, consider an initial pure state \(\ket{\psi_i}\) in the single-excitation subspace. Conditioned on no photon emission during the time interval \([0,t]\), the unnormalized state evolves as
\begin{equation}
    \ket{\tilde{\psi}(t)}=\mathrm e^{-\mathrm i\mathcal H_{\mathrm{eff}}t/\hbar}\ket{\psi_i}.
\end{equation}
The no-jump probability is given by the squared norm
\begin{equation}
    P_{\mathrm{no}}(t)=\braket{\tilde{\psi}(t)|\tilde{\psi}(t)}=\bra{\psi_i}\mathrm e^{\mathrm i\mathcal H_{\mathrm{eff}}^\dagger t/\hbar}\mathrm e^{-\mathrm i\mathcal H_{\mathrm{eff}}t/\hbar}\ket{\psi_i}.
\end{equation}
After averaging over the no-jump trajectory and all possible jump trajectories, the density matrix becomes
\begin{equation}
    \rho(t)=\mathrm e^{-\mathrm i\mathcal H_{\mathrm{eff}}t/\hbar}\ket{\psi_i}\bra{\psi_i}\mathrm e^{\mathrm i\mathcal H_{\mathrm{eff}}^\dagger t/\hbar}+\left[1-\bra{\psi_i}\mathrm e^{\mathrm i\mathcal H_{\mathrm{eff}}^\dagger t/\hbar}\mathrm e^{-\mathrm i\mathcal H_{\mathrm{eff}}t/\hbar}\ket{\psi_i}\right]\ket{G}\bra{G}.
\end{equation}
This expression is equivalent to the Lindblad evolution in the single-excitation sector. In particular, if the initial state is an eigenstate of \(\mathcal H_{\mathrm{eff}}\),
\begin{equation}
    \mathcal H_{\mathrm{eff}}\ket{\psi_\alpha}=\hbar\left(\omega_\alpha-\frac{\mathrm i}{2}\gamma_\alpha\right)\ket{\psi_\alpha},
\end{equation}
then
\begin{equation}
    \rho(t)=\mathrm e^{-\gamma_\alpha t}\ket{\psi_\alpha}\bra{\psi_\alpha}+\left(1-\mathrm e^{-\gamma_\alpha t}\right)\ket{G}\bra{G}.
\end{equation}
Thus, the real part of the complex eigenvalue gives the collective eigenfrequency \(\omega_\alpha\), while its imaginary part gives the collective radiative decay rate \(\gamma_\alpha\). In the following, the most subradiant state is identified as the eigenstate of \(\mathcal H_{\mathrm{eff}}\) with the smallest \(\gamma_\alpha\).

For an infinite one-dimensional chain, translational symmetry allows the single-excitation eigenmodes to be written as Bloch spin waves,
\begin{equation}
    \ket{k_z}=\sum_j \mathrm e^{\mathrm i k_z z_j}\sigma_j^+\ket{G},
\end{equation}
where $k_z$ is restricted to the first Brillouin zone, $|k_z|\leq \pi/a$. These modes diagonalize the effective non-Hermitian Hamiltonian as
\begin{equation}
    \mathcal H_{\mathrm{eff}}\ket{k_z}=\hbar\left(\omega_{k_z}-\frac{\mathrm i}{2}\gamma_{k_z}\right)\ket{k_z}.
\end{equation}
where $\omega_{k_z}$ is the collective frequency shift. For dipoles polarized parallel to the chain, the collective frequency shift is
\begin{equation}
    \frac{\omega_{k_z}^{\parallel}}{\gamma_0}=-\frac{3}{2k_0^3a^3}\mathrm{Re}\left[\mathrm{Li}_3\!\left(\mathrm e^{\mathrm i(k_0+k_z)a}\right)+\mathrm{Li}_3\!\left(\mathrm e^{\mathrm i(k_0-k_z)a}\right)-\mathrm i k_0a\,\mathrm{Li}_2\!\left(\mathrm e^{\mathrm i(k_0+k_z)a}\right)-\mathrm i k_0a\,\mathrm{Li}_2\!\left(\mathrm e^{\mathrm i(k_0-k_z)a}\right)\right],
\end{equation}
where $\mathrm{Li}_n(x)$ is the polylogarithm. The corresponding collective decay rate is
\begin{equation}
    \frac{\gamma_{k_z}^{\parallel}}{\gamma_0}=\frac{3\pi}{2k_0a}\sum_{g_z}\Theta\!\left(k_0-|k_z+g_z|\right)\left[1-\frac{(k_z+g_z)^2}{k_0^2}\right],
\end{equation}
where $g_z=2\pi m/a$ with $m\in\mathbb Z$ are reciprocal lattice vectors, and $\Theta(x)$ is the Heaviside step function. The step function enforces the light-cone condition. Therefore, for modes satisfying
\begin{equation}
    |k_z+g_z|>k_0\quad\text{for all reciprocal lattice vectors }g_z,
\end{equation}
the collective decay rate vanishes. In particular, for a subwavelength chain with $a<\lambda_0/2$, modes near the Brillouin-zone edge satisfy $|k_z|>k_0$ and are perfectly subradiant in the infinite-chain limit.

\begin{figure}[h]
  \centering
  \includegraphics[width=1\linewidth]{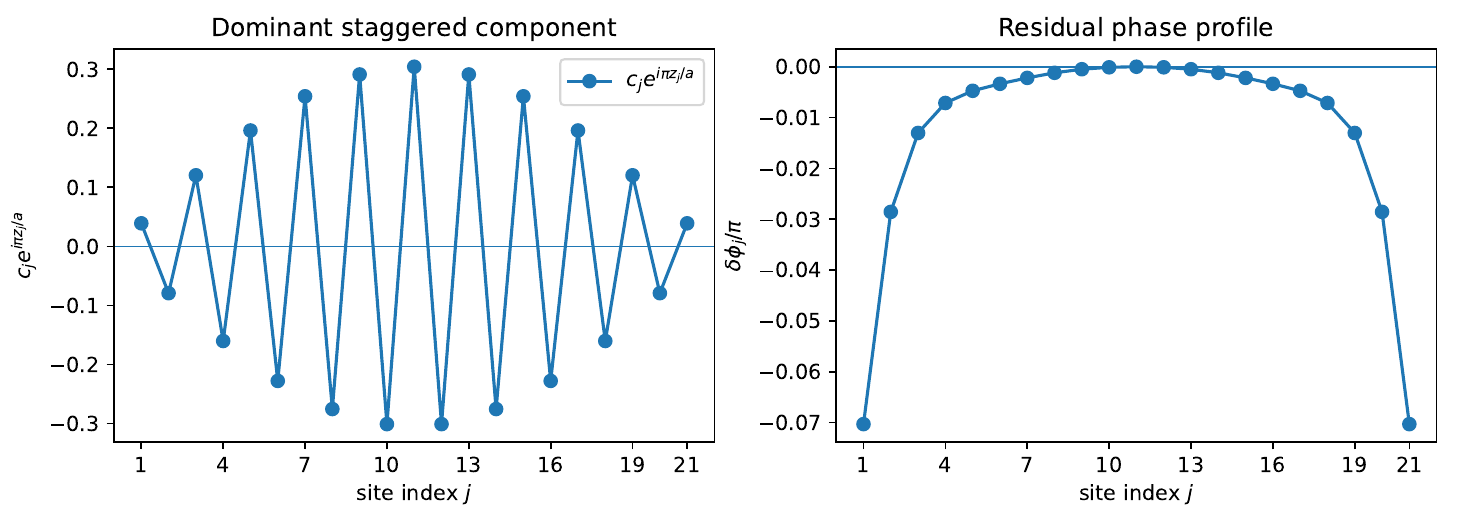}
  \caption{
    Decomposition of the most subradiant eigenstate in a finite atomic chain. 
    Left: dominant staggered component \(c_j e^{i\pi z_j/a}\) of the numerically obtained most subradiant eigenstate for \(N=21\) and \(a=\lambda_0/3\), showing the alternating sign structure and the nonuniform envelope. 
    Right: residual phase profile \(\delta\varphi_j/\pi\) after removing the dominant staggered phase. 
    The residual phase is small compared with the leading \(\pi\) phase difference between neighboring atoms, justifying its neglect in the qualitative phase-matching discussion. 
    In the numerical fidelity calculations, the exact eigenstate including this residual phase profile is used as the target state.
}
  \label{fig:S1_1}
\end{figure}

For a finite chain, the collective eigenmodes are obtained by numerical diagonalization of $\mathcal H_{\mathrm{eff}}$ in the single-excitation subspace. The most subradiant eigenstate can be written as
\begin{equation}
    \ket{\psi_{\mathrm{sub}}}=\sum_{j=1}^{N}c_j\mathrm e^{\mathrm i\pi z_j/a}\text{e}^{\text{i}\delta \varphi_j}\sigma_j^+\ket{G},
\end{equation}
where the staggered phase factor captures the dominant Bloch component near the Brillouin-zone edge, while the coefficients $c_j$ describe the nonuniform envelope. The envelope of the most subradiant state is well approximated by the ansatz
\begin{equation}
    c^{\text{ans}}_j=\sqrt{\frac{2 }{N+1}}\sin\left(\frac{\pi j}{N+1}\right),\;\;j=1,...,N.
\end{equation}
In the numerical calculations presented in the main text, however, the coefficients $c_j$ are obtained directly from the eigenvectors of $\mathcal H_{\mathrm{eff}}$.
Figure~\ref{fig:S1_1} shows this decomposition for a representative chain with \(N=21\) and \(a=\lambda_0/3\). 
The dominant component \(c_j e^{i\pi z_j/a}\) is real and exhibits the expected alternating sign structure together with a nonuniform envelope. 
The residual phase \(\delta\varphi_j\) remains much smaller than the leading \(\pi\) phase difference between neighboring atoms and is mainly visible near the chain edges, where the envelope amplitude is small. 
This supports the approximation used in the main text: \(\delta\varphi_j\) can be neglected in the qualitative phase-matching discussion, while the exact eigenstate including this residual phase profile is retained in all numerical fidelity calculations.

\section{II. Derivation of the scattering operator}
In this section, we derive the scattering operator that describes the interaction between a free-electron wave packet and an atomic chain.

{We start with the single atom case.} The total Hamiltonian can be written as
\begin{equation}
    H=H_{0F}+H_{0B}+H_I,
\end{equation}
where $H_{0F}$ is the free-electron Hamiltonian, $H_{0B}$ is the atomic Hamiltonian, and $H_I$ describes the Coulomb interaction between the free electron and the bound electron in the atom.

The free-electron Hamiltonian is expanded around the central momentum $p_0$ as
\begin{equation}
    H_{0F}=\sqrt{m_e^2c^4+p_z^2c^2}-m_ec^2\approx E_0+v_0(p_z-p_0)+\frac{(p_z-p_0)^2}{2\gamma^3m_e}+\frac{p_{\perp}^2}{2\gamma m_e},
\end{equation}
where $E_0$ and $v_0$ are the central energy and velocity of the electron. In the following derivation, we use the non-recoil approximation: $E_0\gg\hbar\omega_0$~\cite{garciadeabajoCompleteExcitationDiscrete2022}, which means that we ignore the quadratic terms in the dispersion, since we always consider electrons with $v_0>0.01c$. In all parameter regimes considered here, $v_0\leq0.0668c$, corresponding to $\gamma-1<2.3\times10^{-3}$. We therefore set $\gamma\simeq1$. {Additionally, since the interaction strength between electrons and atoms is sufficiently weak, we assume that the transverse evolution of the electron beam is given simply by free-space evolution, that is, the interaction between free electron and atom will only change the longitudinal distribution of the electron wave packet. We call it \textit{free-space evolution approximation of the transverse distribution}.}

The atomic Hamiltonian satisfies
\begin{equation}
    H_{0B}\ket{n}=E_n\ket{n},
\end{equation}
and we subsequently restrict it to the two-level subspace $\{\ket{g},\ket{e}\}$. For a single atom, the Coulomb interaction in the non-retarded approximation is
\begin{equation}
    H_I=\frac{e^2}{4\pi\varepsilon_0}\left[\frac{1}{|\mathbf r_F-\mathbf r_B|}-\frac{1}{|\mathbf r_F-\mathbf r_0|}\right],
\end{equation}
where $\mathbf r_{F}$, $\mathbf r_{B}${, and $\mathbf r_0$} are the position \textit{operators} of the free electron, the bound electron{, and the central nucleus of the atom}, respectively.

With dipole approximation, the Hamiltonian reduces to
\begin{equation}
    H_I=\frac{e^2}{4\pi\varepsilon_0}\frac{\boldsymbol{l}\cdot(\mathbf r_{F}-\mathbf r_0)}{|\mathbf r_{F}-\mathbf r_0|^3}.
\end{equation}
Since the position vectors $\mathbf r_{F}$, $\mathbf r_0$ are operators here, we expand the Hamiltonian in the free electron eigenstates
\begin{equation}
    H_I=\left(\sum_{m'}\ket{m'}\bra{m'}\right)H_I\left(\sum_m\ket{m}\bra{m}\right),
\end{equation}
where we use $\ket{m}$ as an eigenstate of a free-space electron wave packet. Due to the free-space evolution approximation we made, we can choose a fixed transverse profile $\phi_\perp$ for the free electron,
\begin{equation}
    \ket{m}=\ket{\phi_\perp,p_z},
\end{equation}
with 
\begin{equation}
    \braket{\boldsymbol{r}|\phi_\perp,p_z}=
    \frac{\phi_{\perp}(\boldsymbol r_{\perp},z)\text e^{\frac{\text ip_zz}{\hbar}}}{\sqrt{2\pi\hbar}},
\end{equation}
where $\phi_{\perp}(\boldsymbol r_{\perp},z)$ is the transverse distribution of the free electron which also includes the effect of diffraction. Thus, $\sum_m$ will be $\int\mathrm d p_z$. Then the interaction Hamiltonian becomes (for notational simplicity, we henceforth denote \(\ket{\phi_{\perp},p_z}\) by \(\ket{p_z}\))
\begin{equation}
    H_I=v_0\iint\frac{\mathrm d p_z'\mathrm dp_z}{2\pi}\ket{p_z'}\bra{p_z}\mathcal G(p_z'-p_z)\sigma_++\text{h.c.},
\end{equation}
where 
\begin{equation}
    \mathcal G(p_z'-p_z)=\frac{e^2}{4\pi\varepsilon_0\hbar v_0}\int\mathrm d^3\mathbf r\,|\phi_\perp(\mathbf r_\perp,z)|^2\mathrm e^{-\text{i}(p_z'-p_z)z/\hbar}\frac{\boldsymbol l\cdot(\mathbf r-\mathbf r_0)}{|\mathbf r-\mathbf r_0|^3}.
\end{equation}
We then move to the interaction picture. 
\begin{equation}
    H_I^I=v_0\iint\frac{\mathrm d p_z'\mathrm dp_z}{2\pi}\ket{p_z'}\bra{p_z}\mathcal G(p_z'-p_z)\text e^{\text{i}[v_0(p_z'-p_z)+\hbar\omega_0]t/\hbar}\sigma_++\text{h.c.}.
\end{equation}

With the single-atom formalism established, we now generalize it to an array of \(N\) atoms. For $N$ atoms located at $\mathbf{r}_j=z_j\hat{\mathbf z}$, the total Hamiltonian in the interaction picture will be 
\begin{equation}
    H_I^I(t)=v_0\sum_{j=1}^{N}\iint\frac{\mathrm d p_z'\mathrm dp_z}{2\pi}\ket{p_z'}\bra{p_z}\mathcal G_j(p_z'-p_z)\text e^{\text{i}[v_0(p_z'-p_z)+\hbar\omega_0]t/\hbar}\sigma_j^++\text{h.c.},
\end{equation}
with 
\begin{equation}
    \mathcal G_j(p_z'-p_z)=\frac{e^2}{4\pi\varepsilon_0\hbar v_0}\int\mathrm d^3\mathbf r\,|\phi_\perp(\mathbf r_\perp,z)|^2\mathrm e^{-\frac{\text{i}(p_z'-p_z)z}{\hbar}}\frac{\boldsymbol l\cdot(\mathbf r-z_j\hat{z})}{|\mathbf r-z_j\hat{z}|^3},
\end{equation}
is the coupling strength.
Using the first-order Magnus expansion, the scattering operator is 
\begin{equation}
    \begin{split}
        S&=\exp\left(-\frac{\text i}{\hbar}\int_{-\infty}^{+\infty}\mathrm d tH_I^I(t)\right)\\
        &=\exp\left(-\frac{\text i}{\hbar}\int_{-\infty}^{+\infty}\mathrm d t\; v_0\sum_{j=1}^{N}\iint\frac{\mathrm d p_z'\mathrm dp_z}{2\pi}\ket{p_z'}\bra{p_z}\mathcal G_j(p_z'-p_z)\text e^{\text{i}[v_0(p_z'-p_z)+\hbar\omega_0]t/\hbar}\sigma_j^++\text{h.c.}\right)\\
        &=\exp\left(-\frac{\text i v_0}{\hbar}\sum_{j=1}^{N}\iint\frac{\mathrm d p_z'\mathrm dp_z}{2\pi}\ket{p_z'}\bra{p_z}\mathcal G_j(p_z'-p_z)\Big[2\pi\hbar\delta[v_0(p_z'-p_z)+\hbar\omega_0]\Big]\sigma_j^++\text{h.c.}\right)\\
        &=\exp\left(-\frac{\text i v_0}{\hbar}\sum_{j=1}^{N}\iint{\mathrm d p_z'\mathrm dp_z}\ket{p_z-\tfrac{\hbar\omega_0}{v_0}}\bra{p_z}\mathcal G_j(p_z'-p_z)\Big[\hbar\delta[v_0(p_z'-p_z)+\hbar\omega_0]\Big]\sigma_j^++\text{h.c.}\right)\\
        &=\exp\left[-\text{i}\left(\int\mathrm d p_z\ket{p_z-\tfrac{\hbar\omega_0}{v_0}}\bra{p_z}\sum_{j=1}^{N}\mathcal G_j\left(-\tfrac{\hbar\omega_0}{v_0}\right)\sigma_j^++\mathrm{h.c.}\right)\right]\\
    \end{split}
\end{equation}
With $b=\int\mathrm d p_z\ket{p_z-\tfrac{\hbar\omega_0}{v_0}}\bra{p_z}$ and
\begin{equation}
    g_j=\mathcal G_j\left(-\tfrac{\hbar\omega_0}{v_0}\right)\text{e}^{-\text{i}\frac{\omega_0}{v_0}z_j}=\frac{e^2}{4\pi\varepsilon_0\hbar v_0}\int\mathrm d^3\mathbf r\,|\phi_\perp(\mathbf r_\perp,z)|^2\mathrm e^{\text{i}\frac{\omega_0}{v_0}(z-z_j)}\frac{\boldsymbol l\cdot(\mathbf r-z_j\hat{z})}{|\mathbf r-z_j\hat{z}|^3},
\end{equation} 
we obtain
\begin{equation}
    S=\exp\left[-\text{i}\left(gbS_++\mathrm{h.c.}\right)\right].
\end{equation}
where 
\begin{equation}
    S_+=\frac{1}{g}\sum_jg_j\text{e}^{\text{i}\frac{\omega_0}{v_0}z_j}\sigma_j^+,
\end{equation}
with $g=\sqrt{\sum_j|g_j|^2}$.

\section{III. Diffraction-free and well-focused beam and corresponding fidelity}
{
We start with a diffraction-free, well-focused straight electron trajectory, and the transverse distribution can be approximated as
\begin{equation}
    |\phi_{\perp}(\boldsymbol{r}_{\perp},z)|^2
    =\delta^{(2)}(\boldsymbol{r}_{\perp}-\boldsymbol{r}_{\perp0}),
\end{equation}
as commonly assumed in similar free-electron--bound-electron interaction problems~\cite{goverFreeElectronBoundElectronResonant2020,zhaoQuantumEntanglementModulation2021,ruimyAtomicResolutionQuantumMeasurements2021}. 
The scattering matrix then becomes
\begin{equation}
    S=\exp\left[-\mathrm{i}\left(gb S_++\text{h.c.}\right)\right],
\end{equation}
with
\begin{equation}
    S_+=\frac{1 }{\sqrt{N}}\sum_{j}\mathrm{e}^{\mathrm{i}\omega_0z_j/v_0}\sigma^+_j.
\end{equation}}

Therefore, the conditional fidelity is determined by the overlap between a uniformly weighted pseudo-spin wave and the nonuniform envelope \(c_j\),
\begin{equation}
    F_N=\left|\frac{1}{\sqrt{N}}\sum_{j=1}^{N}c_j\right|^2 .
\end{equation}
Using the sine-envelope ansatz for the most subradiant state,
\begin{equation}
    c_j\simeq\sqrt{\frac{2}{N+1}}
    \sin\left(\frac{\pi j}{N+1}\right),
\end{equation}
we obtain
\begin{equation}
    F_N=
    \frac{2}{N(N+1)}
    \cot^2\left[\frac{\pi}{2(N+1)}\right].
\end{equation}
In the large-\(N\) limit, this expression approaches
\begin{equation}
    \lim_{N\rightarrow\infty}F_N=\frac{8}{\pi^2}.
\end{equation}
This gives the asymptotic fidelity limit shown in Fig.~{2(c)} in the main text. The result reflects the fact that a straight trajectory can match the staggered phase, but not the nonuniform envelope of the most subradiant state.

\section{IV. Gaussian electron beam and parabolic trajectory}
In this section, we extend the analysis to a more general case where the transverse profile of the electron beam cannot be approximated by a delta function. We assume that the free electron has a Gaussian transverse wave function. Ignoring phase factors that do not enter the probability density, we write
\begin{equation}
    \phi(\boldsymbol r_\perp,z)=\sqrt{\frac{2}{\pi}}\frac{1}{w(z)}\exp\left[-\frac{|\boldsymbol r_\perp-\boldsymbol R(z)|^2}{w^2(z)}\right]\frac{\text e^{\frac{\text i p_z z}{\hbar}}}{\sqrt{2\pi\hbar}},
\end{equation}
where
\begin{equation}
    w(z)=w_0\sqrt{1+\left(\frac{z}{z_R}\right)^2},\qquad z_R=\frac{\pi w_0^2}{\lambda_e}.
\end{equation}
Here \(w(z)\) characterizes the free diffraction of the electron wave packet.

In the presence of a uniform transverse DC electric field \(\mathbf E=E_{\mathrm{DC}}\hat{\boldsymbol x}\), the wave-packet center follows a parabolic trajectory with transverse profile remaining Gaussian~\cite{kudlisEvolutionTwistedElectron2026}. Since the coupling coefficient depends only on the transverse probability density, the field-induced phase factor does not affect the interaction considered here. We therefore take
\begin{equation}
    \boldsymbol R(z)=R(z)\hat{\boldsymbol x},\qquad R(z)=r_0+\alpha\left(\frac{z}{a}\right)^2,
\end{equation}
with
\begin{equation}
    \alpha=\frac{eE_{\mathrm{DC}}a^2}{2m_ev_0^2}.
\end{equation}
The transverse probability density is then
\begin{equation}
    |\phi_\perp(\boldsymbol r_\perp,z)|^2=\frac{2}{\pi w^2(z)}\exp\left[-\frac{2|\boldsymbol r_\perp-\boldsymbol R(z)|^2}{w^2(z)}\right].
\end{equation}

The coupling strength to the \(j\)th atom is
\begin{equation}
    g_j=\frac{e^2}{4\pi\varepsilon_0\hbar v_0}\int \mathrm dz\int \mathrm d^2\boldsymbol r_\perp\,|\phi_\perp(\boldsymbol r_\perp,z)|^2\mathrm e^{\text i\frac{\omega_0}{v_0}(z-z_j)}\frac{\boldsymbol l\cdot[\boldsymbol r_\perp+(z-z_j)\hat{\boldsymbol z}]}{\left[r_\perp^2+(z-z_j)^2\right]^{3/2}}.
\end{equation}
Substituting the Gaussian transverse distribution gives
\begin{equation}
    \begin{split}
        g_j&=\frac{e^2}{4\pi\varepsilon_0\hbar v_0}\int \mathrm dz\,\mathrm e^{\text i\frac{\omega_0}{v_0}(z-z_j)}\frac{2}{\pi w^2(z)}\int_0^\infty r\,\mathrm dr\int_0^{2\pi}\mathrm d\theta\,\exp\left[-\frac{2\left(r^2+R^2(z)-2rR(z)\cos\theta\right)}{w^2(z)}\right]\\
        &\quad\times\frac{l_xr\cos\theta+l_yr\sin\theta+l_z(z-z_j)}{\left[r^2+(z-z_j)^2\right]^{3/2}}.
    \end{split}
\end{equation}
The angular integral can be carried out analytically. Since
\begin{equation}
    \int_0^{2\pi}\mathrm d\theta\,\mathrm e^{A\cos\theta}=2\pi I_0(A),\qquad \int_0^{2\pi}\mathrm d\theta\,\cos\theta\,\mathrm e^{A\cos\theta}=2\pi I_1(A),\qquad \int_0^{2\pi}\mathrm d\theta\,\sin\theta\,\mathrm e^{A\cos\theta}=0,
\end{equation}
with
\begin{equation}
    A=\frac{4rR(z)}{w^2(z)},
\end{equation}
we obtain
\begin{equation}
    \begin{split}
        g_j&=\frac{e^2}{\pi\varepsilon_0\hbar v_0}\int \mathrm dz\,\mathrm e^{\text i\frac{\omega_0}{v_0}(z-z_j)}\frac{\mathrm e^{-\frac{2R^2(z)}{w^2(z)}}}{w^2(z)}\int_0^\infty r\,\mathrm dr\,\mathrm e^{-\frac{2r^2}{w^2(z)}}\frac{l_z(z-z_j)I_0\!\left(\frac{4rR(z)}{w^2(z)}\right)+l_xrI_1\!\left(\frac{4rR(z)}{w^2(z)}\right)}{\left[r^2+(z-z_j)^2\right]^{3/2}}.
    \end{split}
\end{equation}
This expression is the Gaussian-beam coupling coefficient \(g_j\). The coupling to the \(j\)th atom is not determined solely by the transverse wave function at \(z=z_j\); instead, it receives contributions from the whole longitudinal wave packet, weighted by the Coulomb kernel centered at \(z_j\) and by the phase factor \(\mathrm e^{\text i\omega_0 (z-z_j)/v_0}\). Equivalently, using \(Z=z-z_j\), one may write
\begin{equation}
    \begin{split}
        g_j&=\frac{e^2}{\pi\varepsilon_0\hbar v_0}\int_{-\infty}^{+\infty}\mathrm dZ\,\mathrm e^{\text i\frac{\omega_0}{v_0}Z}\frac{\mathrm e^{-\frac{2R^2(Z+z_j)}{w^2(Z+z_j)}}}{w^2(Z+z_j)}\int_0^\infty r\,\mathrm dr\,\mathrm e^{-\frac{2r^2}{w^2(Z+z_j)}} \frac{l_zZI_0\!\left(\frac{4rR(Z+z_j)}{w^2(Z+z_j)}\right)+l_xrI_1\!\left(\frac{4rR(Z+z_j)}{w^2(Z+z_j)}\right)}{\left(r^2+Z^2\right)^{3/2}}.
    \end{split}
\end{equation}

For a dipole polarized along the chain, \(\boldsymbol l=l_z\hat{\boldsymbol z}\), this reduces to
\begin{equation}
    g_j=\frac{e^2l_z}{\pi\varepsilon_0\hbar v_0}\int \mathrm dz\,\mathrm e^{\text i\frac{\omega_0}{v_0}(z-z_j)}\frac{\mathrm e^{-\frac{2R^2(z)}{w^2(z)}}}{w^2(z)}\int_0^\infty r\,\mathrm dr\,\mathrm e^{-\frac{2r^2}{w^2(z)}}\frac{(z-z_j)I_0\!\left(\frac{4rR(z)}{w^2(z)}\right)}{\left[r^2+(z-z_j)^2\right]^{3/2}}.
\end{equation}
For a dipole polarized along \(\hat{\boldsymbol x}\), one obtains
\begin{equation}
    g_j=\frac{e^2l_x}{\pi\varepsilon_0\hbar v_0}\int \mathrm dz\,\mathrm e^{\text i\frac{\omega_0}{v_0}(z-z_j)}\frac{\mathrm e^{-\frac{2R^2(z)}{w^2(z)}}}{w^2(z)}\int_0^\infty r\,\mathrm dr\,\mathrm e^{-\frac{2r^2}{w^2(z)}}\frac{rI_1\!\left(\frac{4rR(z)}{w^2(z)}\right)}{\left[r^2+(z-z_j)^2\right]^{3/2}}.
\end{equation}
For a dipole polarized along \(\hat{\boldsymbol y}\), i.e., $\boldsymbol l=l_y\hat{\boldsymbol y}$, the coupling vanishes in this mirror-symmetric geometry
\begin{equation}
    g_j=0,
\end{equation}
because the angular integral over the \(\sin\theta\) term is zero. The vanishing of the coupling for a \(y\)-oriented dipole is consistent with the well-focused, non-diffracting limit.

\section{V. Representative feasible parameters for different \(a\) and \(N\)}

In the main text, we focus on representative examples to illustrate the mechanism of free-electron writing of the most subradiant state. 
Here we provide additional numerical checks showing that the same mechanism remains feasible over a broader range of lattice constants and chain lengths. 
For each pair of \(a/\lambda_0\) and \(N\), we optimize the electron-beam parameters appearing in Sec.~IV, including the closest approach \(r_0\), the beam waist \(w_0\), the transverse DC electric field \(E_{\mathrm{DC}}\), and the phase-matching order \(n\). 
The electron velocity is chosen according to
\begin{equation}
    \frac{\omega_0}{v_0}=\frac{(2n+1)\pi}{a},
\end{equation}
so that the longitudinal momentum transfer is matched to the Brillouin-zone edge modulo a reciprocal lattice vector.

The optimization is performed subject to the three feasibility criteria (which are also satisfied in the main text):
\begin{equation}
    F\geq 0.995,\qquad
    P_{\mathrm{exc}}\geq 10^{-6},\qquad
    E_{\mathrm{DC}}\leq 10^8~\mathrm{V/m}.
\end{equation}
Here \(F\) is the conditional fidelity between the heralded one-excitation state and the exact most subradiant eigenstate of the finite chain, \(P_{\mathrm{exc}}\) is the one-loss excitation probability, and \(E_{\mathrm{DC}}\) is the transverse static electric field required to generate the parabolic trajectory. 
For the parameters listed below, \(E_{\mathrm{DC}}\) is obtained from the trajectory curvature in Sec.~IV through
\begin{equation}
    E_{\mathrm{DC}}=\frac{2m_ev_0^2\alpha}{ea^2}.
\end{equation}
Using the phase-matching condition, this can also be written as
\begin{equation}
    E_{\mathrm{DC}}=
    \frac{8m_ec^2}{e\lambda_0^2}
    \frac{\alpha}{(2n+1)^2},
\end{equation}
where \(\lambda_0=780~\mathrm{nm}\) is used for the numerical values in Table~\ref{tab:feasible_parameters}.

Table~\ref{tab:feasible_parameters} gives representative parameter sets obtained from the numerical optimization. 
For each pair of \(a/\lambda_0\) and \(N\), we find at least one set of beam parameters satisfying the fidelity, heralding-probability, and DC field constraints. 
The entries are not unique; nearby choices in the beam-parameter space can give comparable performance. 
The table therefore serves as a numerical check that the protocol remains feasible over a range of atomic spacings and number of atoms.

\begin{table}[h]
\centering
\begin{tabular}{c c c c c c c c}
\hline
\(a/\lambda_0\) & \(N\) & \(n\) & \(r_0~(\mathrm{nm})\) & \(w_0~(\mathrm{nm})\) & \(F~(\%)\) & \(P_{\mathrm{exc}}~(10^{-6})\) & \(E_{\mathrm{DC}}~(10^7~\mathrm{V/m})\) \\
\hline
\(1/20\) & 20  & 1 & 10.00 & 4.83 & 99.59 & 18.17 & 4.25 \\
\(1/20\) & 40  & 1 & 10.00 & 1.27 & 99.53 & 25.78 & 2.08 \\
\(1/20\) & 60  & 1 & 10.00 & 1.37 & 99.54 & 38.68 & 1.28 \\
\(1/20\) & 80  & 1 & 10.00 & 1.76 & 99.53 & 55.41 & 0.71 \\
\(1/20\) & 100 & 1 & 10.00 & 2.20 & 99.55 & 68.14 & 0.47 \\
\hline
\(1/10\) & 20  & 1 & 10.00 & 2.11 & 99.57 & 16.55 & 8.16 \\
\(1/10\) & 40  & 1 & 15.00 & 6.64 & 99.51 & 7.95 & 2.20 \\
\(1/10\) & 60  & 2 & 15.00 & 1.79 & 99.58 & 4.84 & 0.59 \\
\(1/10\) & 80  & 2 & 15.00 & 2.05 & 99.59 & 6.67 & 0.40 \\
\(1/10\) & 100 & 2 & 15.00 & 2.56 & 99.55 & 8.10 & 0.26 \\
\hline
\(1/5\) & 20  & 3 & 15.00 & 3.92 & 99.57 & 3.36 & 1.35 \\
\(1/5\) & 40  & 3 & 15.00 & 4.09 & 99.51 & 6.37 & 0.42 \\
\(1/5\) & 60  & 4 & 18.00 & 3.07 & 99.58 & 1.79 & 0.20 \\
\(1/5\) & 80  & 4 & 18.00 & 3.95 & 99.51 & 2.46 & 0.12 \\
\(1/5\) & 100 & 4 & 20.00 & 4.60 & 99.51 & 1.41 & 0.09 \\
\hline
\end{tabular}
\caption{
Representative parameter sets for different lattice constants \(a/\lambda_0\) and chain lengths \(N\). 
The columns \(F\), \(P_{\mathrm{exc}}\), and \(E_{\mathrm{DC}}\) give the conditional fidelity, the excitation probability, and the required transverse DC electric field, respectively. 
The fidelity \(F\) is given in percent, \(P_{\mathrm{exc}}\) is given in units of \(10^{-6}\), and \(E_{\mathrm{DC}}\) is given in units of \(10^7~\mathrm{V/m}\). The displayed values of \(w_0\) and \(E_{\mathrm{DC}}\) are rounded to two decimal places; the reported \(F\) and \(P_{\mathrm{exc}}\) are calculated using the unrounded beam parameters.
}
\label{tab:feasible_parameters}
\end{table}

It can be seen from the table that the choice of \(w_0\) does not follow an obvious trend. 
By contrast, \(r_0\) and \(n\) tend to increase as \(a\) and \(N\) increase. 
For a fixed lattice spacing \(a\), a longer chain generally requires a weaker DC electric field.

\section{VI. Effect of the incident electron energy spread}

We consider an incident electron with a Gaussian kinetic-energy distribution
\begin{equation}
    P(E)=\frac{1}{\sqrt{2\pi}\sigma_E}\text e^{-\frac{(E-E_0)^2}{2\sigma_E^2}},
\end{equation}
where
\begin{equation}
    \sigma_E=\frac{\Delta E_{\text{FWHM}}}{2\sqrt{2\ln2}},
\end{equation}
and \(E_0\) is the central kinetic energy. An energy component \(E\) has velocity
\begin{equation}
    v(E)=c\sqrt{1-\left(1+\frac{E}{m_ec^2}\right)^{-2}},
\end{equation}
and imprints the longitudinal wave vector
\begin{equation}
    q(E)=\frac{\omega_0}{v(E)}.
\end{equation}

For each energy component, the coupling coefficients \(g_j(E)\) are evaluated using the Gaussian-beam model in Sec.~IV, with the same experimental parameters \(w_0\), \(r_0\), and \(E_{\text{DC}}\) as those optimized for the central energy \(E_0\). The energy dependence of the electron diffraction and trajectory is included through \(v(E)\). The atomic one-excitation state associated with the loss of one energy quantum \(\hbar\omega_0\) is
\begin{equation}
    \ket{\Phi(E)}=\sum_{j=1}^{N}g_j(E)\text e^{\text i q(E)z_j}\sigma_j^+\ket{G}.
\end{equation}
The corresponding excitation probability is
\begin{equation}
    P_{\text{exc}}(E)=\braket{\Phi(E)|\Phi(E)}=\sum_{j=1}^{N}|g_j(E)|^2.
\end{equation}

The one-loss component of the joint electron--atom state can be written as
\begin{equation}
    \ket{\Psi_1}=\int\mathrm dE\,\sqrt{P(E)}\ket{E-\hbar\omega_0}\ket{\Phi(E)}.
\end{equation}
After tracing over the outgoing electron energy, the unnormalized conditional atomic density matrix is
\begin{equation}
    \rho_{\text{at}}^{(\Delta E_{\text{FWHM}})}=\int\mathrm dE\,P(E)\ket{\Phi(E)}\bra{\Phi(E)}.
\end{equation}
The heralding probability is
\begin{equation}
    P_{\text{herald}}^{(\Delta E_{\text{FWHM}})}=\mathrm{Tr}\rho_{\text{at}}^{(\Delta E_{\text{FWHM}})}=\int\mathrm dE\,P(E)P_{\text{exc}}(E),
\end{equation}
and the conditional fidelity with the target most subradiant state is
\begin{equation}
    F_{\Delta E_{\text{FWHM}}}=\frac{\bra{\psi_{\text{sub}}}\rho_{\text{at}}^{(\Delta E_{\text{FWHM}})}\ket{\psi_{\text{sub}}}}{P_{\text{herald}}^{(\Delta E_{\text{FWHM}})}}=\frac{\int\mathrm dE\,P(E)\left|\braket{\psi_{\text{sub}}|\Phi(E)}\right|^2}{\int\mathrm dE\,P(E)P_{\text{exc}}(E)}.
\end{equation}

The physical effect of the energy spread can be seen directly from the matrix elements of the conditional atomic state,
\begin{equation}
    \left[\rho_{\text{at}}^{(\Delta E_{\text{FWHM}})}\right]_{jl}=\int\mathrm dE\,P(E)g_j(E)g_l^*(E)\text e^{\text i q(E)(z_j-z_l)}.
\end{equation}
A finite energy spread produces a distribution of the imprinted wave vector \(q(E)\). Different energy components therefore write collective states with slightly different relative phases along the atomic chain. The phase difference accumulated between two atoms at \(z_j\) and \(z_l\) is
\begin{equation}
    \delta\varphi_{jl}(E)=\left[q(E)-q(E_0)\right](z_j-z_l).
\end{equation}
The phase variation increases with the separation \(|z_j-z_l|\), making the long-range coherence of a longer chain more sensitive to the incident energy spread.

We numerically evaluate the energy integrals. For every parameter set, \(w_0\), \(r_0\), \(E_{\text{DC}}\), and \(n\) are fixed at their monochromatic optimized values listed in Table~\ref{tab:feasible_parameters}. Table~\ref{tab:finite_width} shows the conditional fidelity for \(\Delta E_{\text{FWHM}}=0\), \(0.1~\mathrm{eV}\), \(0.25~\mathrm{eV}\), and \(0.6~\mathrm{eV}\).

\begin{table}[h]
\centering
\begin{tabular}{c c c c c c c}
\hline
\(a/\lambda_0\) & \(N\) & \(n\) & \(F_0~(\%)\) & \(F_{0.1~\mathrm{eV}}~(\%)\) & \(F_{0.25~\mathrm{eV}}~(\%)\) & \(F_{0.6~\mathrm{eV}}~(\%)\) \\
\hline
\(1/20\) & 20  & 1 & 99.59261 & 99.59190 & 99.58805 & 99.56647 \\
\(1/20\) & 40  & 1 & 99.52945 & 99.52723 & 99.51557 & 99.44952 \\
\(1/20\) & 60  & 1 & 99.54223 & 99.53866 & 99.51992 & 99.41376 \\
\(1/20\) & 80  & 1 & 99.52606 & 99.51983 & 99.48715 & 99.30215 \\
\(1/20\) & 100 & 1 & 99.54802 & 99.53793 & 99.48501 & 99.18386 \\
\hline
\(1/10\) & 20  & 1 & 99.57130 & 99.57125 & 99.57103 & 99.56973 \\
\(1/10\) & 40  & 1 & 99.51156 & 99.51140 & 99.51053 & 99.50561 \\
\(1/10\) & 60  & 2 & 99.58117 & 99.57512 & 99.54336 & 99.36370 \\
\(1/10\) & 80  & 2 & 99.59136 & 99.58155 & 99.53002 & 99.23162 \\
\(1/10\) & 100 & 2 & 99.55388 & 99.53781 & 99.45111 & 98.95001 \\
\hline
\(1/5\) & 20  & 3 & 99.56826 & 99.56782 & 99.56553 & 99.55253 \\
\(1/5\) & 40  & 3 & 99.50666 & 99.50503 & 99.49646 & 99.44796 \\
\(1/5\) & 60  & 4 & 99.57518 & 99.56026 & 99.47928 & 99.01570 \\
\(1/5\) & 80  & 4 & 99.51106 & 99.48201 & 99.33176 & 98.50231 \\
\(1/5\) & 100 & 4 & 99.50575 & 99.45845 & 99.22222 & 97.89375 \\
\hline
\end{tabular}
\caption{
Conditional fidelity for Gaussian incident electron energy distributions with different FWHM energy spreads. \(F_0\) denotes the monochromatic result. The electron-beam and trajectory parameters are fixed at their monochromatic optimized values.
}
\label{tab:finite_width}
\end{table}

The excitation probability is highly robust against the incident energy spread. For all parameter sets in Table~\ref{tab:finite_width}, the relative change in \(P_{\text{herald}}^{(\Delta E_{\text{FWHM}})}\) remains below \(4\times10^{-4}\), even at \(\Delta E_{\text{FWHM}}=0.6~\mathrm{eV}\). This confirms that the energy dependence of the coupling amplitudes has a negligible effect on the total excitation probability.

The fidelity reduction is governed by the phase variation accumulated across the chain. For \(\Delta E_{\text{FWHM}}=0.1~\mathrm{eV}\), the additional fidelity reduction remains below \(5\times10^{-4}\) for all parameter sets. At \(\Delta E_{\text{FWHM}}=0.6~\mathrm{eV}\), the reduction remains small for short chains and becomes appreciable for longer chains. The largest reduction is \(1.612\times10^{-2}\), obtained for \(a/\lambda_0=1/5\) and \(N=100\). The numerical results also exhibit the expected approximate quadratic dependence on the energy spread.

An incident energy spread on the order of \(0.6~\mathrm{eV}\) is compatible with high-fidelity preparation in shorter chains. For longer chains, monochromation to approximately \(0.1~\mathrm{eV}\) suppresses the additional finite-energy-width correction below \(5\times10^{-4}\). Such an incident energy spread can be achieved using a gun energy filter or electron monochromator~\cite{tromp2023gun}.

\section{VII. Two-dipole interference factor}

Here we derive the interference factor for two identical radiating dipoles separated by a distance $d$. This factor is used in the main text as a simple guide for understanding the additional decay-rate suppression in the antisymmetric two-chain state.

Consider two identical two-level emitters located at $\mathbf r_1$ and $\mathbf r_2$, with separation vector $\mathbf r=\mathbf r_1-\mathbf r_2$ and $r=d$. Their collective radiative decay is determined by the dissipative coupling matrix
\begin{equation}
    \Gamma_{ij}=\frac{2\mu_0\omega_0^2}{\hbar}\mathbf d_i^*\cdot\mathrm{Im}\,\mathbf G(\mathbf r_i,\mathbf r_j,\omega_0)\cdot\mathbf d_j.
\end{equation}
For identical dipoles with $\mathbf d_1=\mathbf d_2=\mathbf d$, the diagonal elements are $\Gamma_{11}=\Gamma_{22}=\gamma_0$, while the off-diagonal element $\Gamma_{12}$ describes the radiative interference between the two dipoles.

The symmetric and antisymmetric single-excitation states are
\begin{equation}
    \ket{\psi_{\pm}}=\frac{1}{\sqrt{2}}\left(\sigma_1^+\pm\sigma_2^+\right)\ket{g_1g_2}.
\end{equation}
Their decay rates are
\begin{equation}
    \gamma_{\pm}=\gamma_0\pm\Gamma_{12}.
\end{equation}
Thus, the antisymmetric state is subradiant when $\Gamma_{12}>0$, and the symmetric state is subradiant when $\Gamma_{12}<0$. In general, the smaller decay rate is
\begin{equation}
    \gamma_{\mathrm{sub}}=\gamma_0-|\Gamma_{12}|.
\end{equation}

We now evaluate $\Gamma_{12}$. The imaginary part of the free-space dyadic Green's tensor gives the standard result
\begin{equation}
    \frac{\Gamma_{12}}{\gamma_0}=\frac{3}{2}\left[\left(1-\left|\hat{\mathbf d}\cdot\hat{\mathbf r}\right|^2\right)\frac{\sin k_0d}{k_0d}+\left(1-3\left|\hat{\mathbf d}\cdot\hat{\mathbf r}\right|^2\right)\left(\frac{\cos k_0d}{(k_0d)^2}-\frac{\sin k_0d}{(k_0d)^3}\right)\right],
\end{equation}
where $k_0=\omega_0/c$, $\hat{\mathbf d}=\mathbf d/|\mathbf d|$, and $\hat{\mathbf r}=\mathbf r/d$.

For dipoles perpendicular to the separation direction, $\hat{\mathbf d}\cdot\hat{\mathbf r}=0$, this reduces to
\begin{equation}
    \frac{\Gamma_{12}^{\perp}}{\gamma_0}=\frac{3}{2}\left[\frac{\sin k_0d}{k_0d}+\frac{\cos k_0d}{(k_0d)^2}-\frac{\sin k_0d}{(k_0d)^3}\right].
\end{equation}
Therefore, the normalized minimum decay rate of the two-dipole system is
\begin{equation}
    \frac{\gamma_{\mathrm{sub}}^{\perp}}{\gamma_0}=1-\frac{3}{2}\left|\frac{\sin k_0d}{k_0d}+\frac{\cos k_0d}{(k_0d)^2}-\frac{\sin k_0d}{(k_0d)^3}\right|.
\end{equation}

In the small-distance limit $k_0d\ll1$, defining $x=k_0d$, we have
\begin{equation}
    \frac{\sin x}{x}+\frac{\cos x}{x^2}-\frac{\sin x}{x^3}=\frac{2}{3}-\frac{2x^2}{15}+O(x^4).
\end{equation}
Thus,
\begin{equation}
    \frac{\gamma_{\mathrm{sub}}^{\perp}}{\gamma_0}=1-\frac{3}{2}\left(\frac{2}{3}-\frac{2(k_0d)^2}{15}\right)+O((k_0d)^4)=\frac{(k_0d)^2}{5}+O((k_0d)^4).
\end{equation}
This gives the quadratic suppression of the most subradiant two-dipole state at small separation.

\section{VIII. Representative optimization for two parallel chains}

We consider two identical parallel chains, denoted by \(L\) and \(R\), each containing \(N_{\mathrm c}=10\) atoms, corresponding to a total atom number \(N=2N_{\mathrm c}=20\). The chains are parallel to the \(z\) axis and separated along \(x\) by a distance \(d\), with atomic positions
\begin{equation}
    \mathbf r_{Lj}=-\frac{d}{2}\hat{\mathbf x}+z_j\hat{\mathbf z},
    \qquad
    \mathbf r_{Rj}=+\frac{d}{2}\hat{\mathbf x}+z_j\hat{\mathbf z}.
\end{equation}
The two electron-path components are centered above the corresponding chains and follow
\begin{equation}
    \mathbf r_{\mu,\mathrm e}(z)
    =x_\mu\hat{\mathbf x}
    +\left[r_0+\alpha\left(\frac{z}{a}\right)^2\right]\hat{\mathbf y}
    +z\hat{\mathbf z},
    \qquad
    x_L=-\frac{d}{2},\quad x_R=+\frac{d}{2}.
\end{equation}

We denote by \(G_{\mu\nu,j}\) the full complex coupling coefficient between electron path \(\mu\in\{L,R\}\) and the \(j\)th atom in chain \(\nu\in\{L,R\}\), evaluated using the Gaussian-beam coupling integral derived in Sec.~IV. The coefficients \(G_{LL,j}\) and \(G_{RR,j}\) describe the couplings to the chain directly below each electron path, while \(G_{LR,j}\) and \(G_{RL,j}\) describe electron-path crosstalk with the neighboring chain. All four coupling channels are retained in the numerical calculation.

After recombination at the antisymmetric output port, the unnormalized heralded atomic state is
\begin{equation}
    \ket{\widetilde{\psi}_{-}}
    =
    \frac{1}{\sqrt{2}}
    \sum_{j=1}^{N_{\mathrm c}}
    \left[
    \left(G_{LL,j}-G_{RL,j}\right)\sigma_{Lj}^{+}
    +
    \left(G_{LR,j}-G_{RR,j}\right)\sigma_{Rj}^{+}
    \right]
    \ket{G_LG_R}.
\end{equation}
The corresponding excitation probability and normalized heralded state are
\begin{equation}
    P_{\mathrm{exc}}
    =
    \braket{\widetilde{\psi}_{-}|\widetilde{\psi}_{-}},
    \qquad
    \ket{\psi_{-}}
    =
    \frac{\ket{\widetilde{\psi}_{-}}}{\sqrt{P_{\mathrm{exc}}}}.
\end{equation}

The target state used in the optimization is
\begin{equation}
    \ket{\psi_A^{\mathrm{tar}}}
    =
    \frac{1}{\sqrt{2}}
    \left(
    \ket{\psi_{\mathrm{sub}}}_L\ket{G_R}
    -
    \ket{G_L}\ket{\psi_{\mathrm{sub}}}_R
    \right),
\end{equation}
where \(\ket{\psi_{\mathrm{sub}}}_{L,R}\) denotes the exact most subradiant eigenstate of an isolated \(N_{\mathrm c}\)-atom chain. The conditional fidelity is evaluated as
\begin{equation}
    F=\left|\braket{\psi_A^{\mathrm{tar}}|\psi_{-}}\right|^2.
\end{equation}

For the representative calculation, we take
\(\lambda_0=780~\mathrm{nm}\),
\(a=\lambda_0/10=78~\mathrm{nm}\),
\(d=a\), and \(r_0=5~\mathrm{nm}\).
The longitudinal momentum transfer is chosen as
\begin{equation}
    \frac{\omega_0/v_0}{\pi/a}=2.994,
\end{equation}
corresponding to \(v_0\simeq0.0668c\). This slight deviation from exact odd-integer phase matching partially compensates for the distortion induced by electron-path crosstalk.

\begin{figure}[h]
  \centering
  \includegraphics[width=0.95\linewidth]{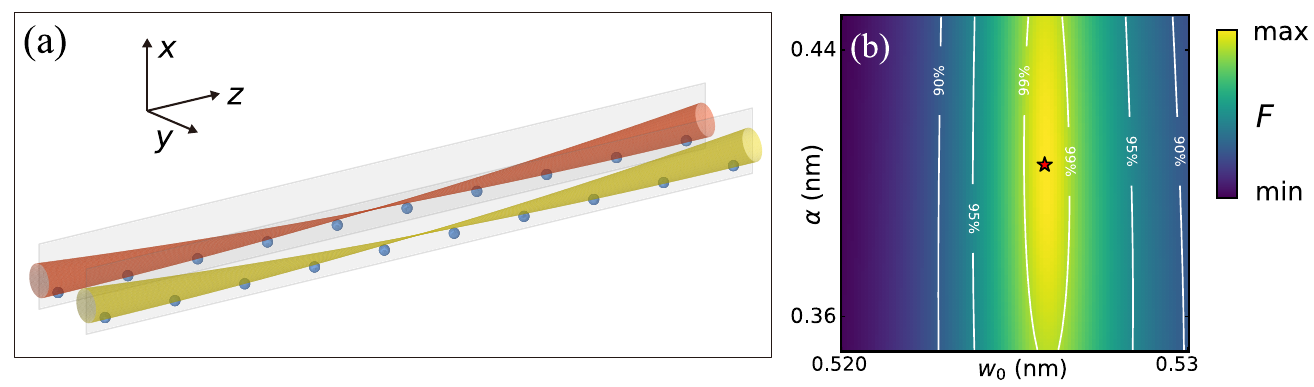}
  \caption{
    Representative optimization for the two-parallel-chain configuration including electron-path crosstalk.
    (a) Geometry of two Gaussian electron-path components interacting with two parallel atomic chains. The orange and yellow tubes denote the two electron-path components, while the blue spheres denote the atoms.
    (b) Conditional fidelity \(F\) as a function of the Gaussian beam waist \(w_0\) and trajectory curvature \(\alpha\). The white curves are fidelity contours, and the red star marks the maximum-fidelity point.
  }
  \label{fig:S_two_chain_opt}
\end{figure}

Figure~\ref{fig:S_two_chain_opt} shows the numerical scan over
\(0.520~\mathrm{nm}\leq w_0\leq0.530~\mathrm{nm}\) and
\(0.35~\mathrm{nm}\leq\alpha\leq0.45~\mathrm{nm}\).
A direct evaluation at
\((w_0,\alpha)=(0.526~\mathrm{nm},0.403~\mathrm{nm})\)
gives
\begin{equation}
    F=99.37\%,
    \qquad
    P_{\mathrm{exc}}=3.94\times10^{-5}.
\end{equation}

\clearpage
\bibliographystyle{apsrev4-2}
\bibliography{main}